\documentclass[aps,prd,reprint,showpacs,article,floatfix,nofootinbib]{revtex4-1}
\pdfoutput=1

\usepackage{fullpage}
\usepackage{amsmath}
\usepackage{amssymb}
\usepackage{graphicx}
\allowdisplaybreaks[4]

\usepackage{subcaption}

\begin{document}

\title{Specific Heats for Rotating  
Quantum BTZ Black Holes in Extended Thermodynamics}

\author{Roberto Nazario}

\affiliation{Department of Physics and Astronomy, University of
Southern California,
 Los Angeles, CA 90089-0484, U.S.A.}

\begin{abstract}
 In the framework of extended thermodynamics, where the cosmological constant $\Lambda$ plays the role of a dynamical pressure $p$, its conjugate variable $V$ arises naturally. This makes it possible to define $C_p$ and $C_V$, the heat capacities at constant pressure and volume, respectively. We extend our previous work on the heat capacities of the static ``quantum" version of the BTZ black hole defined on a braneworld model to the case where the black hole is rotating. The extra degree of freedom that rotation grants the system imparts it with infinite families of both $C_p$ and $C_V$. We find exact formulae for these heat capacities as functions of the three dimensionless parameters of the theory, and explore some special cases in detail. In all cases considered, at least two physically realizable branches were observed, including both positive and negative heat capacities, signaling both stable and unstable black holes, respectively. Though the critical point seen in the static case disappears, other interesting points arise where the heat capacities diverge. Finally, we discuss the conjectured connection in the literature between the super-entropicity of a black hole and its instability, though much like in the static case, the exact relationship, if any, remains unclear.
\end{abstract}

\maketitle

\section{Introduction}
Classical physics consigns black holes to always absorb, but never emit, radiation, though the picture changes when quantum effects are taken into account. Quantum physics makes the semi-classical physics of black holes compatible with the laws of thermodynamics \cite{Bekenstein:1973ur,Bekenstein:1974ax,Hawking:1975vcx,Hawking:1976de}, imparting a temperature $T{=}\kappa/2\pi$ and an entropy $S{=}A/4$ to them\footnote{We work in units where $G=k_B=c=\hbar=1$.}, where~$\kappa$ is the surface gravity of the event horizon and $A$ is the black hole horizon's surface area \cite{Bardeen:1973gs}. Consequently, a set of laws of black hole thermodynamics arise, the first being, when rotation is included, $dM{=}TdS{+}\Omega dJ$, where $M$ is the black hole's mass, playing the role of  internal energy, $U$, and $\Omega$ and $J$ are the angular velocity and angular momentum, respectively.

With the preceding thermodynamic functions, it becomes possible to calculate the heat capacity. Famously, in asymptotically flat space, the $D{=}4$ Schwarzschild black hole (for which $\Omega{=}J{=}0$) has a temperature given by $T{=}1/8\pi M$, and an entropy $S{=}\pi r_{+}^2{=}4\pi M^2$, where $r_{+}{=}2M$ is the horizon's radius. The black hole's specific heat capacity is thus calculated to be $C(T){=}{-}1/8\pi T^2$ ~\cite{Hawking:1982dh}. Since $C(T){<}0$, the black hole is unstable.

The situation changes when considering asymptotically anti-de Sitter (AdS) black hole solutions. The AdS spacetime possesses a negative cosmological constant, $\Lambda$, which plays the role of  a positive dynamical pressure given by $p{=}{-}\Lambda/8\pi$. Note that in $D$ spacetime dimensions, $\Lambda$ defines a natural length scale of the AdS geometry {\it via} $\Lambda{=}{-}(D{-}1)(D{-}2)/2 L^2$. The presence of a pressure $p$ suggests a natural extension of the first law of black hole thermodynamics to include a pressure-volume work term. Of course, it is not clear \emph{a priori} that the pressure is dynamical at all. Earlier work addressed these concerns and offered some alternative approaches \cite{Henneaux:1984ji,Teitelboim:1985dp,Henneaux:1989zc}.

The aforementioned extension to black hole thermodynamics results in some differences with the black hole thermodynamics discussed at the open. For starters, $M$ no longer corresponds to the black hole's energy, but rather, its enthalpy $H$ \cite{Kastor:2009wy}. Recall that $H{\equiv}U{+}pV$, where $V$ is the volume. The first law thus changes to $dM{=}TdS+Vdp+\Omega dJ$. The  temperature is given in the usual way by $T{=}\partial M/\partial S|_{p,J}$, and it is then possible to define the thermodynamic volume as $V{=}\partial M/\partial p|_{S,J}$. We note here that in various simple  cases $V$ does indeed correspond to a physical spacetime volume, but in general the thermodynamic volume need not coincide with the geometric volume, the former simply being some nontrivial function of the parameters present in the theory ~\cite{Cvetic:2010jb}.

Two heat capacities of considerable interest are $C_p$ and $C_V$, the heat capacities at constant pressure and constant volume, respectively. They are defined as as $C_p(T){\equiv}\partial H/{\partial}T|_{p}$ and $C_V(T){\equiv}\partial U/\partial T|_{V}$. For Schwarzschild-AdS black holes, there exist two families of solutions~\cite{Hawking:1982dh} corresponding to black holes that are large or small compared to the AdS length scale $L$. The small black holes have negative~$C_p$, just as their asymptotically flat counterparts do. The large black holes possess a \emph{positive}~$C_p$ and are thus stable\footnote{Their stability makes them very useful for studying phases of the holographically dual thermal Yang-Mills theory ~\cite{Witten:1998zw}.}. On the other hand, for both large and small Schwarzschild-AdS black holes, $C_V(T){=}0$, which follows~\cite{Dolan:2010ha} from the fact that for Schwarzschild black holes both the entropy and the volume are functions of one another (through $r_+$) and hence are not independent. Specifically, holding $V$ fixed also holds $S$ fixed, and since $C_V$ can equivalently be defined as $C_V(T){\equiv}T\partial S/\partial T|_{V}$, the conclusion follows.

Research on the extended thermodynamics of black hole systems has uncovered quite fortuitous analogies between these gravitational systems and much more familiar systems composed of baryonic matter. The result is that some of these gravitational systems display properties of the Van der Waals fluid, such as phase transitions and their associated critical exponents, triple points, reentrant phase transitions, etc. For a review, see ref. \cite{Kubiznak:2016qmn}. Despite these striking similarities, there are also stark differences between these two classes of systems, mainly in the functional form of their heat capacities. We can glean information about the hidden degrees of freedom from the behavior of the heat capacities, and thus they serve as an important, albeit coarse-grained, diagnostic tool to understand these underlying degrees of freedom and how they differ in gravitational systems compared to typical matter systems. Many of the simplest black hole systems in extended thermodynamics have $C_V{=}0$, and thus we conclude that they possess zero degrees of freedom which can be excited at constant $V$. The Reissner-Nordtr\"{o}m-AdS black hole, for example, has $C_V{=}0$, and yet the system is known to possess a Van der Waals-like phase structure \cite{Chamblin:1998pz,Kubiznak:2014zwa,Johnson:2019vqf}). In the case of Kerr-AdS, we infer from the presence of Schottky peaks in $C_V(T)$~\cite{Johnson:2019vqf} that there exists only a finite window of energy excitations available to the system, and as the temperature rises, these states become saturated and $C_V$ tends to zero. On the other hand, $C_p$ increases monotonically for these black holes beyond their critical point. In the more complex Kerr-Newman case, the Van der Waals phase structure also arises \cite{Dolan:2011xt}.

An important $D{=}3$ asymptotically AdS geometry is the Ba\~{n}ados, Teitelboim, and Zanelli (BTZ) black hole~\cite{Banados:1992wn,Banados:1992gq}. We wish to consider the rotating version of the black hole, in preparation for the rest of this Paper. Since the black hole is rotating, it possesses an angular momentum which modifies some of the thermodynamic quantities compared to the static case. The thermodynamic quantities of interest to us are

\begin{equation}
    \label{BTZtherm1}
    M=\frac{r_{+}^2}{8L^2}+\frac{J^2}{32 r_{+}^2},\;\;T=\frac{r_{+}}{2 \pi L^2}-\frac{J^2}{8\pi r_{+}^3},
\end{equation}

\begin{equation}
    \label{BTZtherm2}
    S=\frac{\pi}{2}r_{+},\;\;p=\frac{1}{8 \pi L^2},\;\;\mathrm{and}\;\;V=\pi r_{+}^2.
\end{equation}

It is possible to eliminate the geometric parameters $L$ and $r_{+}$ between (\ref{BTZtherm1}) and (\ref{BTZtherm2}), resulting in \cite{Frassino:2015oca}

\begin{equation}
    \label{cBTZmasstemp}
    M=\frac{4pS^2}{\pi}+\frac{\pi^2 J^2}{128 S^2},\;\; T=\frac{8pS}{\pi}-\frac{\pi^2 J^2}{64 S^3},
\end{equation}

\begin{equation}
    \label{cBTZvol}
    \mathrm{and}\;\;V=\frac{4S^2}{\pi},
\end{equation}

\noindent from which it follows that

\begin{equation}
    \label{cBTZheatcapacity}
    C_p(p,S)=S\frac{512pS^4-\pi^3 J^2}{512pS^4+3\pi ^3 J^2}\;\; \mathrm{and}\;\; C_V=0.
\end{equation}

First, note that it is possible to solve the $T$ equation in (\ref{cBTZmasstemp}) for $S{=}S(p,T)$, then substitute that into $C_p(p,S)$ to give $C_p(p,T)$ explicitly. However, the result is clunky and not particularly illuminating, thus we will not take that route. Further, note that the form of (\ref{cBTZheatcapacity}) suggests that $C_p{<}0$ when $J^2{>}512pS^4/\pi^3$. This, however, does not signal an instability\cite{Akbar:2011qw} because these values of $J$ are unattainable, as they would result in $T{<}0$ as well. Unfortunately, it seems that the rotating, uncharged BTZ black hole does not lend itself to discussions of instability\footnote{The static, electrically charged version of the classical BTZ does, however, as it has $C_p{>}0$ while $C_V{<}0$. See ref. \cite{Johnson:2019mdp} for details.}. However, we raise the issue because the quantum version of this black hole possesses heat capacities with much more complex and nontrivial functional forms, and it again becomes relevant to consider these instabilities and how they may relate to the relative signs of the heat capacities and to the super-entropicity of the black hole.

The argument was made in ref. \cite{Johnson:2019mdp}  that when considering the stability of black holes, the sign of {\it both} $C_p$ and $C_V$ must be taken into account. Even with a positive $C_p$ a system can find itself with a negative $C_V$ in extended thermodynamics, thereby possessing an instability when $V$ is held fixed. The situation is complicated for the ``quantum" BTZ (qBTZ) because both $C_p$ and $C_V$ possess positive and negative branches in different ranges of the temperature. For the qBTZ, it was observed that for some temperatures both are positive, for others, both are negative, and for yet others, they have opposite signs \cite{Johnson:2023dtf}. The rotating qBTZ, with the extra degree of freedom its rotation affords it, also has a much richer set of heat capacities, and can thus provide many more examples to test any potential relationship between the relative signs of the heat capacities and the black hole's super-entropicity.

It was shown in \cite{Johnson:2019mdp} that for the charged, static BTZ, the heat capacity at constant volume can be written in the form

\begin{equation}
 \label{eq:superentropicCV}
     C_V=-S\left(\frac{1-{\cal R}^2}{3 - {\cal R}^2}\right)\ ,
 \end{equation}

\noindent where 

\begin{equation}
 \label{eq:isoperimeter}
     {\cal R}\equiv\left[\frac{\pi V}{4S^2}\right]^\frac12
 \end{equation}

\noindent for $D{=}3$. The isoperimetric inequality states that $\cal{R}{\geq} \text{1} $, with the inequality corresponding to subentropic systems, and the equality holding for maximally entropic systems. When the inequality is violated and $\cal{R}{<} \text{1} $, the system is said to be superentropic, meaning that the black hole contains more entropy per unit volume than an AdS Schwarzschild black hole of equal mass \cite{Cvetic:2010jb}. Clearly, $C_V{<}0$ when $\cal{R}{<} \text{1} $ for the BTZ, while it turns out that $C_p{>}0$, and thus ref. \cite{Johnson:2019mdp} conjectured that super-entropic systems with $\cal{R}{<}\text{1}$ may always have $C_V{<}0$, when $C_p{>}0$. The hope was that the connection between $C_V$ and the internal microscopic degrees of freedom would be related to the value of $\cal{R}$ in a fairly straightforward way. In fact, ref. \cite{Johnson:2019wcq} found microscopic evidence (using a dual CFT$_2$ language) for such an instability in the case of the classical charged BTZ, as well as generalized exotic BTZ black holes.

Subsequent work has shown that the situation is fairly more nuanced. Ref.~\cite{Cong:2019bud} argued that there is a family of generalized BTZ black holes which exhibit super-entropicity (\emph{i.e.} have ${\cal R}{<}1$) while $C_V{>}0$. Ref. \cite{Johnson:2019wcq} later showed that using a definition of super-entropicity that defines ${\cal R}$ in terms of entropy (instead of area) gives a result that does not contradict the conjecture.  Since then, further stable, yet super-entropic black hole examples have been found (see {\it e.g.,} refs.\cite{Song:2023zre,Jing:2020sdf,JahaniPoshteh:2021clv,Song:2023kvq,He:2023tiz}). However, ref.~\cite{Appels:2019vow} has questioned whether the exotic types of solutions being used can truly be classified as super-entropic, or whether they merely belong to a class of ``critical" black holes, referring to black holes with conical deficits on one of the poles of their horizons, which typically corresponds to the ultra-spinning case. More work is still needed to understand the exact relationship between all the pieces of the puzzle.

Keeping the above in mind, finding new examples of non-trivial behavior for $C_p$ and $C_V$ could thus be of potential value. In this Paper we present computations and observations about the specific heat capacities of the rotating qBTZ black hole which arises from computing the backreaction in a braneworld model \cite{Emparan:1999fd,Emparan:2020znc,Karch:2000ct}. The extended thermodynamics of the static version of this system have also been discussed in recent works\cite{Frassino:2022zaz,Johnson:2023dtf,Frassino:2023wpc}. Many of the features observed there will be present in the rotating case, but new features emerge as well due to the rotation.

Much like in the Schwarzschild-AdS and static qBTZ cases, the rotating qBTZ possesses different branches corresponding to different classes of behavior of black hole solutions. Furthermore, both the static and the rotating qBTZ possess a quantum analog of the Hawking-Page transition. However, unlike Schwarzschild-AdS, we will observe that both $C_p(T)$ and $C_V(T)$ possess negative and positive branches in certain ranges of the temperature, a feature inherited by the rotating qBTZ. Though we have not been able to examine all regions of parameter space exactly (due to the lack of closed-form heat capacities written explicitly as functions of $T$ and $V$), we explore several regions and discuss some instructive features. Surprisingly, a critical point found in the static case disappears when rotation is present, though the numerical plots suggest the presence of other similar points of interest.

\section{The Rotating Quantum BTZ}

\subsection{Review and observations}

It is possible to construct a rotating qBTZ black hole and calculate its thermodynamic functions $M$, $T$, $S$, $p$, $V$, $J$, and $\Omega$, to all orders of the backreaction, by employing a holographic braneworld model. A Karch-Randall AdS$_3$ brane is embedded in an AdS$_4$ bulk C-metric at the location of the regulator surface, and the result is that the \emph{classical }AdS$_4$ black hole induces a ``quantum" black hole in the lower-dimensional brane. See refs. \cite{Emparan:1999fd,Emparan:2020znc,Karch:2000ct} for details.

The brane's position is parametrized by a parameter $\ell$, which is inversely proportional to both the tension of the brane $\tau$ and the acceleration of the black hole in the C-metric viewpoint \cite{Emparan:2020znc,Frassino:2022zaz}. When $L_4$, the AdS$_4$ length scale, is held fixed, the simple result is that $d\tau{=}d\Lambda_3/8\pi G_3{=}{-}dp$ \cite{Frassino:2022zaz}. Hence, variations in the $D{=}3$ pressure are related to variations in the brane tension with no contributions from other parameters \footnote{We are assuming fixed Newton's constants $G_3$ and $G_4$.}.

We omit the full details of the derivation of $M, T$, $S$, $J$, and $\Omega$ here, though we do outline them briefly in Appendix \ref{appendixMTSJO}. For now, we simply reproduce them:
\begin{widetext}
\begin{eqnarray}
    \label{qM}
    M(\nu,z,\alpha)&=&\frac{\sqrt{1+\nu^2}}{2G_3} \frac{(1-\nu z^3)(z^2(1+\nu z)+\alpha^2(1+4z^2+4(1+\alpha^2)\nu z^3-(1+4\alpha^2)z^4))}{(1+3z^2+2\nu z^3-\alpha^2(1-4\nu z^3+3z^4))^2}\ ,
\\
    \label{qT}
    T(\nu,z,\alpha)&=&\frac{1}{2\pi \ell_3(\nu)} \frac{(z^2(1+\nu z)-\alpha^2(1-2\nu z^3+z^4))(2+3(1+\alpha^2)\nu z-4\alpha^2 z^2+\nu z^3+\alpha^2 \nu z^5)}{z(1+\nu z)(1+\alpha^2(1-z^2))(1+3z^2+2\nu z^3-\alpha^2(1-4\nu z^3+3z^4))},
\\
    \label{qS}
     S(\nu,z,\alpha)&=&\frac{\pi \ell_3(\nu) \sqrt{1+\nu^2}}{G_3}\frac{z(1+\alpha^2(1-z^2))}{1+3z^2+2\nu z^3-\alpha^2(1-4\nu z^3+3z^4)},
\\
    \label{qOmega}
    \Omega(\nu,z,\alpha)&=&\frac{\alpha(1+z^2)\sqrt{(1-\nu z^3)(1+\nu z-\alpha^2 z(z-\nu))}}{\ell_3 (\nu) z(1+\nu z)(1+\alpha^2(1-z^2))},\;\;\;\mathrm{and}
\\
    \label{qJ}
    J(\nu,z,\alpha)&=&\frac{\ell_3(\nu) \sqrt{1+\nu^2}}{G_3}\frac{\alpha z(1+z^2)(1+\alpha^2(1-z^2))\sqrt{(1-\nu z^3)(1+\nu z-\alpha^2 z(z-\nu))}}{(1+3z^2+2\nu z^3-\alpha^2(1-4\nu z^3+3z^4))^2},
\end{eqnarray}
\end{widetext}

\noindent where we have temporarily restored the factors of the $D{=}3$ Newton constant, $G_3$, for clarity\footnote{The chosen fixed values of $L_4$ and $G_4$ determine $G_3$ via $G_3=G_4/2L_4$. Having fixed $L_4=1$, we will fix $G_4=2$ for the sake of having $G_3=1$ in the remainder of this Paper.}.

The preceding quantities contain three dimensionless parameters, $\nu$, $z$, and $\alpha$, defined as 
\begin{equation}
\label{parameters}
    \nu\equiv \frac{\ell}{\ell_3},\;\;\; z\equiv\frac{\ell_3}{x_1 r_{+}},\;\;\;\mathrm{and}\;\;\;\alpha\equiv \frac{ax_1}{\sqrt{-\varkappa}\ell_3},
\end{equation} 

\noindent where $\ell_3$ is the ``bare" AdS$_3$ length scale, without taking any of the backreaction effects into account, and appears in both $L_4$, and $L_3$, the AdS$_3$ length scale with backreaction effects included. The parameter  $x_1$ is the single positive root remaining in a polynomial present in the C-metric after part of the bulk is cut off by the brane, and $r_{+}$ is the location of the event horizon. Further, $\varkappa$ in (\ref{parameters}) is a discrete parameter appearing in the bulk metric which takes on the values of -1, 0, or 1, with each value corresponding to a different spacetime geometry on the brane, though $\varkappa{=}{-}1$ results in the BTZ geometry on the brane and is thus the value relevant to our discussion. The spin parameter $a$ which introduces rotation into the picture is contained in the expression for $\alpha$, and it is $\alpha$ that will characterize the rotation in the remainder of this Paper. The parameters $\nu$ and $z$ typically take on values on the interval $[0,\infty)$, but the situation with $\alpha$ included is more complicated. The latter satisfies the constraint

\begin{equation}
    \label{alphaconstraint}
    \alpha^2 \leq \frac{1+\nu z}{z(z-\nu)},
\end{equation}

\noindent which is analogous to the upper bound on the Kerr spin parameter. This constraint arises from restricting $r_{+}$ to real, nonnegative values. For the complete details on this constraint, see ref. \cite{Emparan:2020znc}. However, note that there appears to be a problem whenever $z{\leq}\nu$. We will address this issue more carefully later in this section, though we briefly remark here that (\ref{alphaconstraint}) is only valid when $z{>}\nu$, otherwise, there is no constraint on the value of $\alpha$. Furthermore, equations~(\ref{qM})-(\ref{qJ}) satisfy $dM{=}TdS+\Omega dJ$, if $\nu$ is held fixed and only $z$ and $\alpha$ are allowed to vary, which directly implies the independence of the pressure $p$ on $z$ and $\alpha$. Note that we have written $\ell_3=\ell_3(\nu)$ in preparation for the discussion to follow.

\begin{figure*}
     \centering
     \begin{subfigure}[h]{0.45\textwidth}
         \centering
         \includegraphics[width=\textwidth]{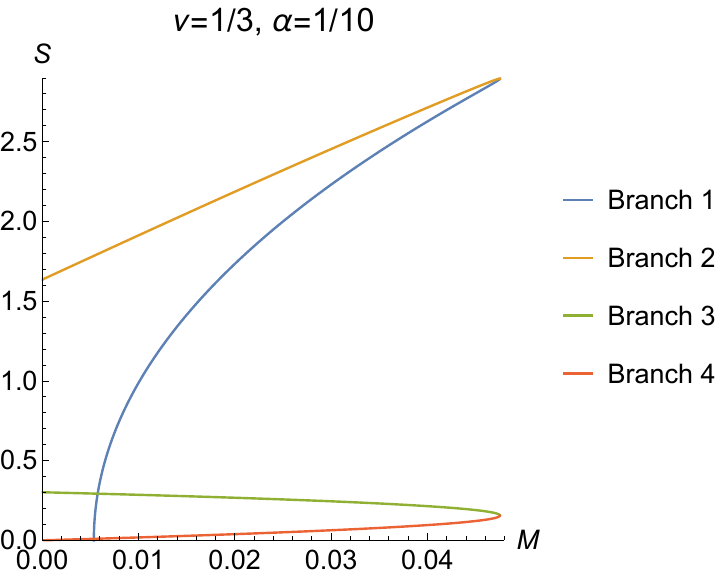}
         \caption{$S$ {\it vs.} $M$ for $\nu=1/3$ and $\alpha$=1/10.}
         \label{fig:SvsMrotating}
     \end{subfigure}
     \hfill
     \begin{subfigure}[h]{0.45\textwidth}
         \centering
         \includegraphics[width=\textwidth]{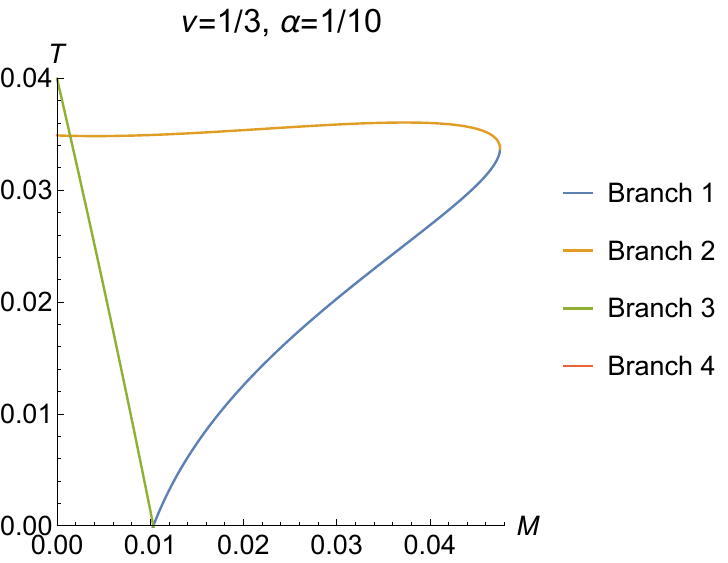}
         \caption{$T$ {\it vs.} $M$ for $\nu=1/3$ and $\alpha=1/10$.}
         \label{fig:TvsMrotating}
         
     \end{subfigure}
     \hfill
      \begin{subfigure}[h]{0.45\textwidth}
         \centering
         \includegraphics[width=\textwidth]{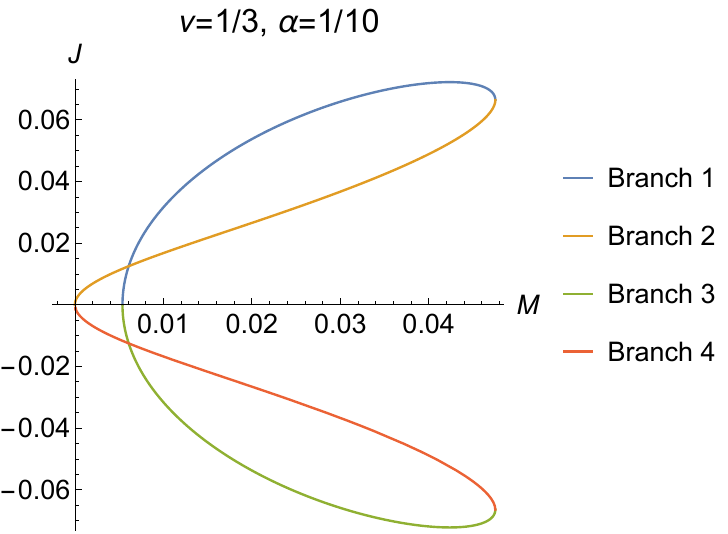}
         \caption{$J$ {\it vs.} $M$ for $\nu=1/3$ and $\alpha$=1/10.}
         \label{fig:JvsMrotating}
     \end{subfigure}
     \begin{subfigure}[h]{0.45\textwidth}
         \centering
         \includegraphics[width=\textwidth]{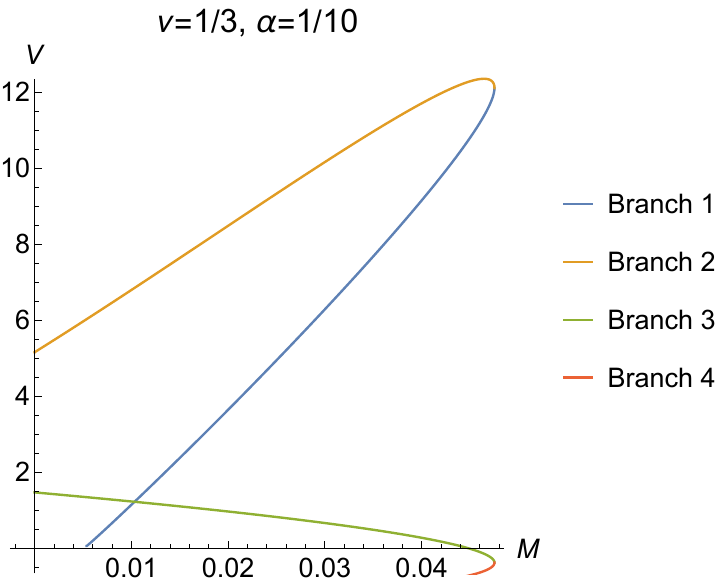}
         \caption{$V$ {\it vs.} $M$ for $\nu=1/3$ and $\alpha=1/10$.}
         \label{fig:VvsMrotating}
     \end{subfigure}
        \caption{Branches corresponding to black holes at fixed $\nu{=}1/3$ and $\alpha{=}1/10$. For positive $M$, there exist four branches, two seen in the static case, and two novel ones in the rotating case. Branch 1 corresponds to stable black holes with positive heat capacity, since both $T$ and $S$ grow monotonically here. Branch 2 contains both positive and negative heat capacity, since $T$ rises and then falls, before rising again, whereas $S$ consistently falls here. Branches 3 and 4 also have positive heat capacity, though not the entirety of both branches may be physically realizable, and are only included for completeness. Note that branch 4 is not shown in figure \ref{fig:TvsMrotating} because it is only present at negative temperatures.}
        \label{fig:vsM}
\end{figure*}

The quantities $\ell, \ell_3, L_4$ and $L_3$ satisfy

\begin{equation}
    \label{L4}
    \frac{1}{L_4^2}=\frac{1}{\ell^2}+\frac{1}{\ell_3^2}\ ,\quad {\rm and} \quad
    \frac{1}{L_3^2}=\frac{2}{L_4^2}\bigg(1-\frac{L_4}{\ell}\bigg)\ .
\end{equation}

Note from the first relation in equations~(\ref{L4}) that fixing $L_4$ imposes a functional dependence between $\ell$ and $\ell_3$, which can be re-expressed via a dependence of $\ell_3$ on $\nu$ given by $\ell_3(\nu)=\sqrt{1+\nu^2}/\nu$. The pressure on the brane is given by $p=1/8\pi L_3^2$, which is given in terms of $\nu$ by

\begin{equation}
    \label{qP}
    p(\nu)=\frac{1+\nu^2-\sqrt{1+\nu^2}}{4\pi \ell_3^2 (\nu) \nu^2}.
\end{equation}

Having the pressure and $\ell_3(\nu)$, one can then seek a thermodynamic volume to complete the First Law, $dM=TdS+Vdp+\Omega dJ$. One can construct this volume directly from the First Law, it is given by

\begin{align}
    \label{generalqV}
    V&=\left({\left.\frac{\partial p}{\partial \nu}\right|_{z,\alpha}}\right)^{-1}\\ & \times \nonumber \left[\left.\frac{\partial M}{\partial \nu}\right|_{z,\alpha} - T\left.\frac{\partial S}{\partial \nu}\right|_{z,\alpha} - \Omega\ \left.\frac{\partial J}{\partial \nu}\right|_{z,\alpha}\right]\ .
\end{align}

\noindent The reader may find the explicit form of $V(\nu,z,\alpha)$ in Appendix \ref{appendixexplicitquantities}.

In the case of static qBTZ, three different branches were observed in the $S$, $T$, and $V$ \emph{vs.} $M$ plots \cite{Johnson:2023dtf,Emparan:2020znc}. Of these branches, there were two for $M{>}0$ and one for $M{<}0$. Further, the mass was found to exist on a finite interval, with a maximum (positive) value and a minimum (negative) value for a given value of $\nu$ \cite{Emparan:2002px,Emparan:1999fd}. The first surprising feature for rotating qBTZ is that for fixed, finite $\nu$ and $\alpha$, two previously unseen branches form for $M{>}0$. The presence of these two new branches will force us to depart from the naming convention we employed in ref. \cite{Johnson:2023dtf}, where we adopted the labeling convention used in ref. \cite{Emparan:2020znc}. We do this simply for the sake of clarity. We shall henceforth simply label the branches as Branch 1, Branch 2, etc., where increasing values of the branch number correspond to increasing values of $z$ which parametrize the branch. See figure \ref{fig:vsM}. Furthermore, we shall exclude the nonphysical branches from our discussion, \emph{i.e.} the portions corresponding to $M{<}0$, $T{<}0$, etc., because unlike in the case of $\alpha{=}0$, the presence of fixed, finite $\alpha{>}0$ results in divergences of the thermodynamic functions for $M{<}0$. The result is that while there is still an upper bound for the mass, the lower bound diverges, and the behavior for $M{<}0$ of $S$, $T$, and $V$ is quite complicated. The two new branches, branches 3 and 4, appear to reach the same maximum mass as branches 1 and 2 in the finite $\alpha$ case for many different fixed values of $\alpha$, though the exact value of this maximum differs from the upper bound in the case where $\alpha{=}0$. It is very difficult to calculate the exact functional dependence on $\nu$ and $\alpha$ of this upper bound, but a numerical check of a wide range of values of the parameters supports our tentative conclusion of the equality of the value of $M$ at the upper bound for all four branches, at least at low $\nu$ and $\alpha$, where interpretation is simplest.

Before we continue discussing the various interesting features of the plots in fig. \ref{fig:vsM}, we must first address a claim alluded to earlier in the Paper, specifically that (\ref{alphaconstraint}) is only valid for $z{>}\nu$, as it will be necessary for the following discussion. The condition in (\ref{alphaconstraint}) sets the upper bound on $\alpha$ for each choice of fixed $\nu$ and $z$. However, at a glance, (\ref{alphaconstraint}) appears to impose the constraint that $z{>}\nu$. Note that as $\nu$ approaches $z$ from below, $\alpha$ diverges, and we thus expect no plots to necessitate truncation in that case, except for to ensure that $T{\geq} 0$. Furthermore, $z{<}\nu$ appears to be forbidden, since it seems to imply purely imaginary $\alpha$. The situation is more nuanced than it appears at first glance. If we refer to the explicit form of $r_{+}$ in ref. \cite{Emparan:2020znc}, given by

\begin{equation}
    \label{rplus}
    r_{+}^2=-\ell^2_3 \varkappa \frac{1+\nu z-\alpha^2 z(z-\nu)}{1-\nu z^3},
\end{equation}

\noindent where the choice $\varkappa{=}\text{sign}(\nu z^3-1)$ was made, we see that when $\nu{=}z$, $r_{+}$ does indeed become independent of $\alpha$, and remains positive. Thus, $\alpha$ can take on any value when $\nu{=}z$ and $r_{+}$ remains real and positive. Further, when $z{<}\nu$ there is again no restriction on $\alpha$. Though it may seem from (\ref{alphaconstraint}) that this region is inaccessible, a more careful look at (\ref{rplus}) reveals that when $z{<}\nu$, $-\alpha^2{\leq} (1+\nu z)/(z(\nu-z))$, which is always true, for any real choice of $\alpha$. We thus conclude that when employing (\ref{alphaconstraint}), we may leave $\alpha$ unconstrained whenever $z{\leq}\nu$, and only impose (\ref{alphaconstraint}) when $z{>}\nu$. This result will truncate some of our heat capacity curves in later parts of the Paper, though we still find novel branches of the heat capacities compared to the static case, as discussed in the preceding section.

In the $S$, $T$, and $V$ \emph{vs.} $M$ plots, as $z$ grows from zero to higher positive values, the black hole is \emph{shrinking} down from infinite size, owing to the fact that $z\propto r_{+}^{-1}$. It is perhaps more instructive to interpret the graphs by considering decreasing values of $z$ instead. For fixed $\nu$ and $\alpha$, the constraint given by (\ref{alphaconstraint}) can be understood to be a cap on the maximum value of $z$, which is our starting point. Smaller black holes are not allowed due to the rotation, reminiscent of the Kerr case. The maximum value of $z$ is given by

\begin{equation}
\label{zmax}
    z_{\emph{max}}=\frac{\nu \left(\alpha^2+1\right)}{2\alpha^2}\left(1+\sqrt{1+\left(\frac{2\alpha}{\nu\left(\alpha^2+1\right)}\right)^2}\right).
\end{equation}

\noindent Note that as $\alpha$ rises sufficiently high, the negative mass branches disappear and only the positive ones remain for small enough values of $\nu$. As $\nu$ also grows, negative mass becomes possible once more,  and a divergence forms as $\nu{\rightarrow}1$. A full analysis of all possible black holes for all values of the parameters is beyond the scope of this work, so we will not expound on this further. See ref. \cite{Emparan:2020znc} for a more complete treatment. Note that for a wide range of the parameters, $z_{\emph{max}}$ is found on Branch 3. Curiously, for both Branch 3 and 1, as $z$ shrinks, so does $M$, though $M$ grows as $z$ falls on Branch 2. Note that while on Branch 3, $M$ falls all the way to zero as $z$ decreases, though on Branch 1 it only decreases to $M(z{=}0){=}\alpha^2\sqrt{1+\nu^2}/2(1-\alpha^2)^2$ as $z$ tends to zero.

We consider still the curves in order of increasing $z$. In the first quadrant, branch 1 joins branch 2 at the maximum mass. When $\alpha{=}0$, branches 1 and 2 join at a cusp for the $S$ \emph{vs.} $M$ plot, a feature very reminiscent of the joining of branches for Schwarzschild-AdS~\cite{Hawking:1982dh}, though for finite $\alpha$ this cusp loses its sharpness and becomes more round the higher $\alpha$ is raised. Branch 2 terminates in the first quadrant on the vertical axis, for positive $S$ (and for $M{=}0$). For increasing $z$, branch 2 connects to an unnumbered ``branch" for negative values of $M$, where it exhibits the complicated behavior alluded to above. After executing this complicated behavior, the curve returns to the $M{>}0$ sector and two more branches arise. Branch 3 connects the $M{<}0$ sector with the $M{>}0$ sector for some $S{>}0$, and branch 4 arises as the curve turns back around at the upper bound of $M$ and terminates at the origin, though much if not all of this branch is usually not available due to (\ref{alphaconstraint}), depending on the exact values of $\nu$ and $\alpha$. Note that in figure \ref{fig:vsM}, two further plots are supplied: $J$ and $\Omega$ \emph{vs.} $M$, where the color-coding of the $S$ \emph{vs.} $M$ branches is applied to all plots, and the branches are broken at the same values of $z$, though it is visually apparent that for arbitrary $\nu$ and $\alpha$, the branch breaks in the different plots don't occur at the exact location of the turning points of $M$. Furthermore, the $J$ and $\Omega$ plots experience discontinuous jumps between Branch 3 and 2, and a closer look reveals that both these functions output complex values when $M{<}0$.

Between $z{=}0$ at the origin and the location where branch 1 joins branch 2, the black hole  shrinks down from infinite size, while its entropy and temperature both rise. It thus has positive specific heat. However, on branch 2, as  $z$ continues to increase and the black hole continues to shrink, the entropy falls, but the temperature's behavior is more complex, exhibiting both rising and falling, depending on the choice of $\nu$ and $\alpha$. Note that an apparent contradiction seems to arise:  for small~$z$, the black hole must be very large, and yet the $V$ {\it vs.} $M$ plot reveals that at small $z$, $V$ is very small, and grows with growing $z$. This is resolved by recalling that the geometric volume built from $r_+$ is distinct from the thermodynamic volume $V$~\cite{Cvetic:2010jb}, and the latter must be in general simply regarded as some function of the the relevant parameters whose precise meaning is not yet entirely understood, but which is likely connected to the underlying degrees of freedom present.

Depending on the values of $\nu$ and $\alpha$, the temperature can rise, then fall, then rise again, flipping the sign of the heat capacity each time. In the static case, we noticed that this behavior was absent at a particular point, given by $(\nu,z){=}(1,1)$, which we argued was the critical point. At the critical point, the local maximum and local minimum of $T$ on Branch 2 joined together into a stationary point. This feature is absent in the rotating case. In ref. \cite{Johnson:2023dtf}, a purely geometric argument was given for why the critical point must be found at the stationary point of the temperature branch. We reproduce the condition here, with the addendum that $\alpha$ also be held fixed during the differentiation, which follows easily from the logic of the original argument (see ref. \cite{Johnson:2023dtf} for the full derivation): 

\begin{equation}
    \label{criticalptcondition}
    \frac{\partial T}{\partial z}\bigg\rvert_{\nu,\alpha}=0,\;\;\frac{\partial^2 T}{\partial z^2}\bigg\rvert_{\nu,\alpha}=0,\;\;\text{and}\;\;\frac{\partial S}{\partial z}\bigg\rvert_{\nu,\alpha} \neq 0.
\end{equation}

\noindent In the case of rotating qBTZ, letting the primes denote partial differentiation with respect to $z$ with $\nu$ and $\alpha$ held fixed, we find that, evaluated at $\nu{=}1$, $z{=}1$, and general $\alpha$, $T'{=} 3\alpha^2/2\sqrt{2}\pi$, $T''{=}(\alpha^2{+}12\alpha^4)/2\sqrt{2}\pi$, and $S'{=}{-}\pi(1{+}2\alpha^2)/3$. Clearly, only $\alpha{=}0$ satisfies (\ref{criticalptcondition}). We thus conclude that this critical point is only present when the black hole does not rotate, and we refer the reader to refs. \cite{Johnson:2023dtf,Frassino:2023wpc} for a much more thorough discussion of the static qBTZ and this interesting critical point.

In preparation for the next section, we remind the reader of the issues raised in the Introduction, where there was a value of $J$ which made $C_p{<}0$ for the classical BTZ, though it was inaccessible because it also made $T{<}0$. It turns out that the phenomenon in the classical BTZ heralds an analogous result in the qBTZ case as well, though in a much more subtle and nuanced manner. The result is not straightforward. Ultimately, what is happening is that there is a competition between the need to satisfy (\ref{alphaconstraint}), while also satisfying that $T{\geq} 0$. We must emphasize that \emph{these two conditions are not equivalent for the quantum BTZ,} as they are in the  classical case. As a result, we observed that over various values of $\nu$ and $\alpha$, sometimes (\ref{alphaconstraint}) was saturated at some $T{>}0$, at which point the branch must terminate lest the horizon be lost and reveal a naked singularity. At other values of $\nu$ and $\alpha$, (\ref{alphaconstraint}) was not yet saturated before $T{=}0$ had been reached, and thus the curve had to terminate to maintain a non-negative temperature, despite the black hole's not rotating maximally. The exact relationship between these two constraints and the black hole solutions they allow is unclear, however, what \emph{is} clear is that even though not the entirety of branches 3 and 4 are physically allowed, at least one of them is \emph{partially} allowed, up to some maximum value of $z$. Thus, we are assured that additional black hole solutions exist in the rotating qBTZ compared with the static case, and we can also therefore expect the heat capacities to exhibit further branches corresponding to these new black holes. In the sections that follow, we will restrict ourselves to plotting portions of the heat capacities where both $T{>}0$ and the constraint given in (\ref{alphaconstraint}) are satisfied, thus ensuring that all the branches plotted are physical.

\begin{figure*}
     \centering
     \begin{subfigure}[h]{0.49\textwidth}
         \centering
         \includegraphics[width=\textwidth]{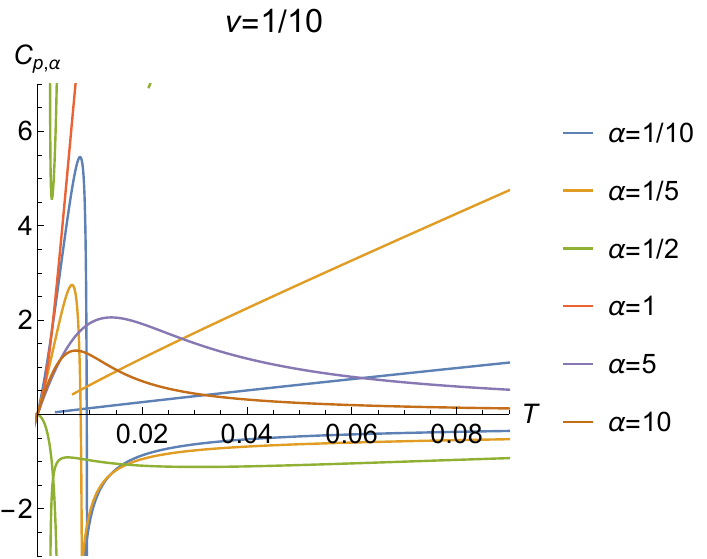}
         \caption{$C_{p,\alpha}$ {\it vs.} $T$ for $\nu=1/10$ and various fixed $\alpha$.}
         \label{fig:Cpplot1}
     \end{subfigure}
     \hfill
     \begin{subfigure}[h]{0.49\textwidth}
         \centering
         \includegraphics[width=\textwidth]{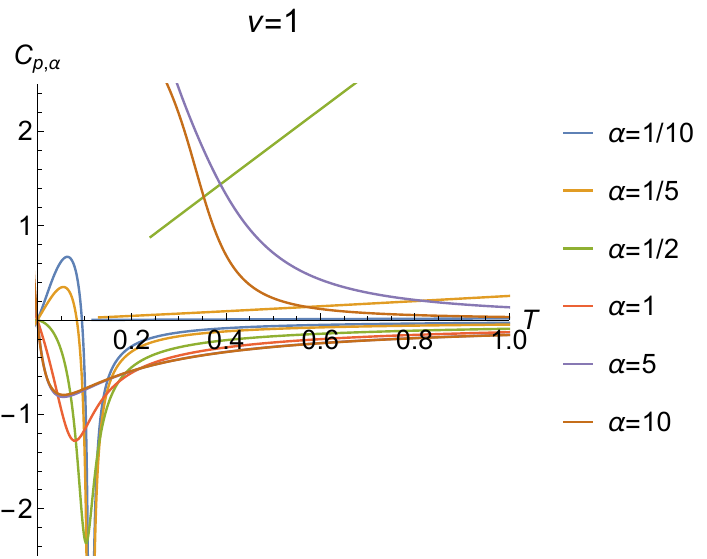}
         \caption{$C_{p,\alpha}$ {\it vs.} $T$ for $\nu=1$ and various fixed $\alpha$.}
         \label{fig:Cpplot2}
         
     \end{subfigure}
     \hfill
      \begin{subfigure}[h]{0.49\textwidth}
         \centering
         \includegraphics[width=\textwidth]{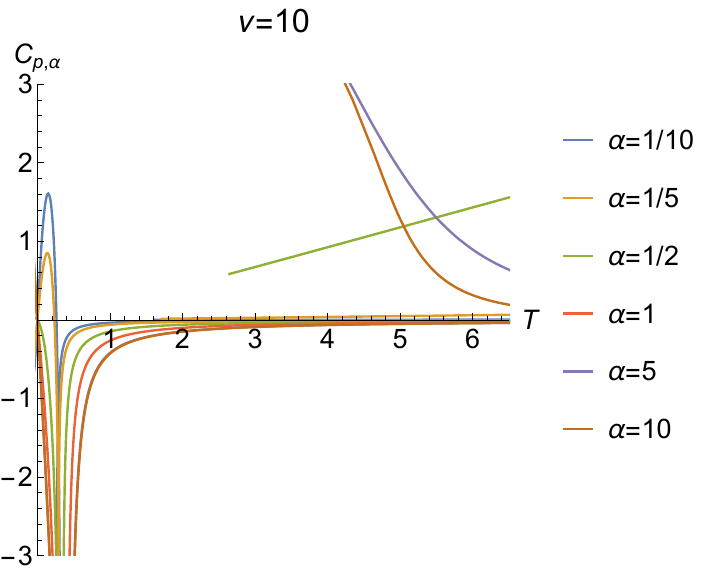}
         \caption{$C_{p,\alpha}$ {\it vs.} $T$ for $\nu=10$ and various fixed $\alpha$.}
         \label{fig:Cpplot3}
     \end{subfigure}
        \caption{$C_{p,\alpha}$ {\it vs.} $T$ for various fixed values of $\nu$ and $\alpha$. Note that curves which terminate suddenly do so because of the upper bound on $z$ imposed by (\ref{alphaconstraint}), and they can be attributed to the new black holes which arise in the case with nonzero rotation. See the text for a more thorough discussion.}
        \label{fig:CpvsT}
\end{figure*}

\subsection{Specific heats at constant pressure}
We turn our attention now to deriving $C_p$ for the rotating qBTZ. The \emph{highly} nontrivial dependence of the thermodynamic functions on $\nu, z$, and $\alpha$ makes it very difficult to find $C_p(T)$ explicitly. We can nevertheless calculate $C_p(\nu,z,\alpha)$, noting that constant $p$ is equivalent to fixed $\nu$. To calculate the heat capacity, we construct it out of $T$, $dT$, and $dS$, where the latter two terms correspond to the variations of the temperature and the entropy, respectively. While there is no unique way to write the general result, one form it takes is

\begin{widetext}
\begin{equation}
    \label{Cpgeneral}
    C_p(\nu,z,\alpha)=T(\nu,z,\alpha) \left[\frac{\partial S}{\partial z}\bigg\rvert_{\nu,\alpha} \frac{dz}{d\lambda}+\frac{\partial S}{\partial \alpha}\bigg\rvert_{\nu,z} \frac{d\alpha}{d\lambda}\right] \left[\frac{\partial T}{\partial z}\bigg\rvert_{\nu,\alpha} \frac{dz}{d\lambda}+\frac{\partial T}{\partial \alpha}\bigg\rvert_{\nu,z} \frac{d\alpha}{d\lambda}\right]^{-1}.
\end{equation}
\end{widetext}

In the derivation of $C_p$, two curious terms arise naturally: $dz/d\lambda$ and $d\alpha/d\lambda$. Their meaning becomes clear if we consider our work in \cite{Johnson:2023dtf} for a moment. In that Paper, $\alpha{=}0$ was fixed, so both $\alpha$ and $\nu$ were fixed. Therefore, the \emph{only} way to vary the temperature and the entropy while keeping $p$ fixed was by varying $z$, which then took on the role of an implicit parameter in a set of parametric equations. With the presence of a new parameter we are free to vary, the situation becomes more complex. Fixing $\nu$ in order to fix $p$ still leaves two parameters free to vary, and the most general situation arises when we can vary them both arbitrarily, so long as we hold $\nu$ fixed. Put another way, there are an infinite number of ways to vary $z$ and $\alpha$ (in order to vary $S$ and $T$) while keeping the pressure constant. We can encapsulate all these infinite ways to vary $z$ and $\alpha$ in a new parameter, $\lambda$, by parametrizing them \emph{via} $z{=}z(\lambda)$ and $\alpha{=}\alpha(\lambda)$. A specific choice of $z(\lambda)$ and $\alpha(\lambda)$ corresponds to picking a specific trajectory in the $z{-}\alpha$ plane (that is, at a fixed pressure slice of the parameter space) along which to calculate the heat capacity. One may even, in principle, eliminate $\lambda$ and find the explicit functional dependence $z{=}z(\alpha)$, or $\alpha{=}\alpha(z)$, depending on one's specific needs. The upshot is that in addition to holding either $z$ or $\alpha$ fixed while the other varies, there are myriad other ways to allow these two parameters to vary, all of which are encapsulated on the chosen curve $z{=}z(\alpha)$, etc. Different curves then result in different heat capacities with novel properties and features which can be studied. In this Paper we will restrict ourselves to considering $C_p$ in the simple cases where either $z$ or $\alpha$ are held fixed and only the remaining one can vary, for simplicity and convenience.

\subsubsection{$C_p$ at fixed $\alpha$}

We begin with the somewhat familiar case of fixed $\alpha$. The case where $\alpha{=}0$ has been worked out as a special case and discussed \cite{Kudoh:2004ub,Johnson:2023dtf,Frassino:2023wpc}, but for general $\alpha$ new features arise in $C_p(T)$. $C_p$ in this case is given by

\begin{equation}
    \label{Cpfixedalpha}
    C_{p,\alpha}(\nu,z,\alpha)=T(\nu,z,\alpha) \frac{\partial S}{\partial z}\bigg\rvert_{\nu,\alpha} \frac{\partial z}{\partial T}\bigg\rvert_{\nu,\alpha}.
\end{equation}

\noindent Readers may refer to Appendix \ref{appendixexplicitquantities} for the explicit form of (\ref{Cpfixedalpha}). We note that at low $\nu$ and $\alpha$ the plots resemble the ones presented in refs. \cite{Johnson:2023dtf,Frassino:2023wpc}, as expected. However, at low $\nu$, as $\alpha$ is increased, the heat capacity goes through a points where it possesses divergences, both to positive and negative infinity, depending on the specific fixed value of $\alpha$. The exact locations of these divergences are difficult to calculate, though they depend on both the choice of $\nu$ and $\alpha$. Beyond this point, as $\alpha$ is raised further, the heat capacity displays a Schottky peak, and it  seems to display this Schottky peak as $\alpha$ is raised without bound, though the value of $C_{p,\alpha}$ at this peak decreases as $\alpha$ increases. See $\alpha{=}5$ and $\alpha{=}10$ in figure \ref{fig:Cpplot1}. The exact locations of the divergences and the formation of the Schottky peak, and whether they are truly present for all $\alpha$ as $\alpha$ is raised without bound is hard to determine exactly because the high order of the polynomials one must solve makes it very difficult to explicitly calculate these features analytically.

Raising $\nu$ and repeating the above observations reveals that at sufficiently high values of $\nu$, the Schottky peak fails to form as one raises $\alpha$. Instead, a divergence forms as $T$ approaches some lower bound from above (see figure \ref{fig:Cpplot3}). For sufficiently high values of $\nu$ and $\alpha$, most branches of the heat capacity become negative for all values of $T$. Of course, one must be cautious when interpreting results for fixed values of $\nu$ that are too high. It is only in the limit of small $\nu$ that the theory of gravity is nearly massless, and for arbitrary $\nu$, where the strength of the backreaction may be significant, it is not entirely clear \emph{a priori} what the appropriate approach is, or which rules still apply.

In all three plots of figure \ref{fig:CpvsT}, there are straight lines which increase monotonically where they are present, but begin abruptly for $T>0$. These heat capacities correspond to black holes arising from branches 3 and 4, the new branches discussed in the previous section. They terminate suddenly because equation (\ref{alphaconstraint}) sets the upper bound on $z$ for their fixed values of $\nu$ and $\alpha$, and it turns out that the upper bound of $z$ does not correspond to $T{=}0$ in those cases, as discussed in the previous section.

\subsubsection{$C_p$ at fixed $z$}
    
The presence of a new parameter allows us to hold $z$ fixed when studying $C_p$. In the case of fixed $z$, the heat capacity is

\begin{equation}
    \label{Cpfixedz}
    C_{p,z}(\nu,z,\alpha)=T(\nu,z,\alpha) \frac{\partial S}{\partial \alpha}\bigg\rvert_{\nu,z} \frac{\partial \alpha}{\partial T}\bigg\rvert_{\nu,z}.
\end{equation}

\noindent Readers may refer to Appendix \ref{appendixexplicitquantities} for the explicit form of (\ref{Cpfixedz}).

\begin{figure*}
     \centering
     \begin{subfigure}[h]{0.40\textwidth}
         \centering
         \includegraphics[width=\textwidth]{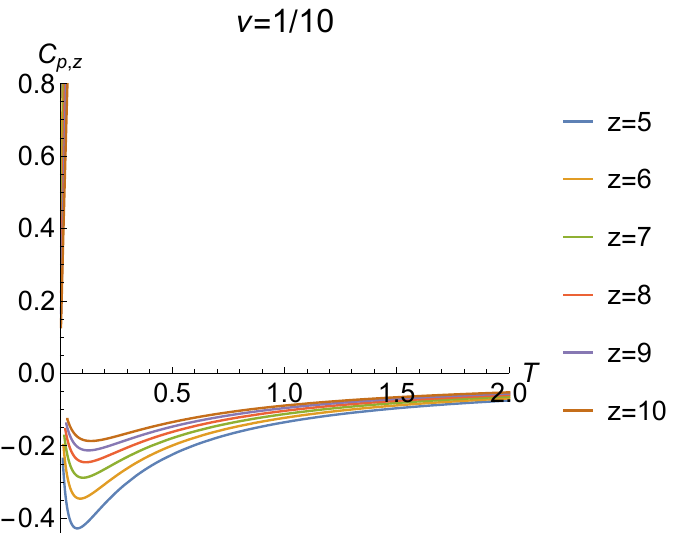}
         \caption{$C_{p,z}$ {\it vs.} $T$ for $\nu=1/10$ and various fixed $z$.}
         \label{fig:Cpplotfixedz1}
     \end{subfigure}
     \hfill
     \begin{subfigure}[h]{0.40\textwidth}
         \centering
         \includegraphics[width=\textwidth]{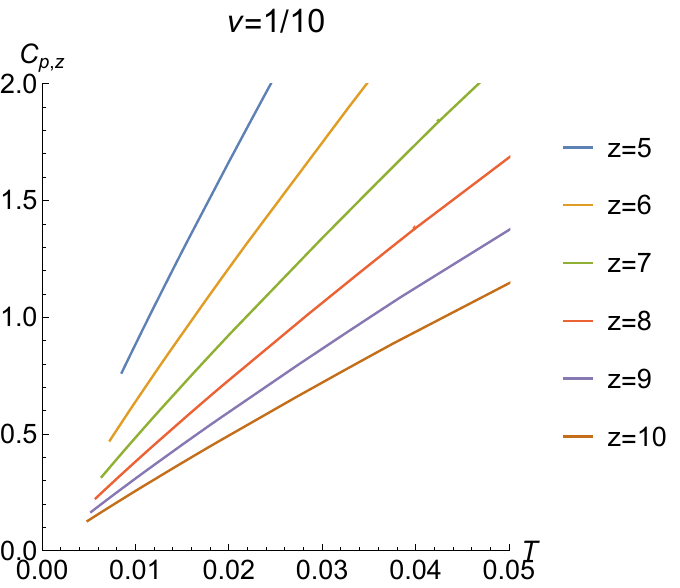}
         \caption{$C_{p,z}$ {\it vs.} $T$ for $\nu=1/10$ and various fixed $z$.}
         \label{fig:Cpplotfizedz2}
     \end{subfigure}
     \hfill
      \begin{subfigure}[h]{0.40\textwidth}
         \centering
         \includegraphics[width=\textwidth]{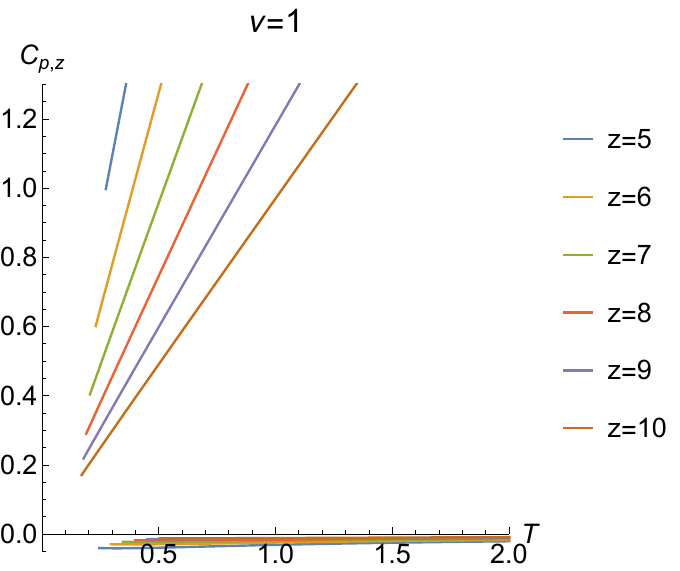}
         \caption{$C_{p,z}$ {\it vs.} $T$ for $\nu=1$ and various fixed $z$.}
         \label{fig:Cpplotfixedz3}
     \end{subfigure}
     \hfill
     \begin{subfigure}[h]{0.40\textwidth}
         \centering
         \includegraphics[width=\textwidth]{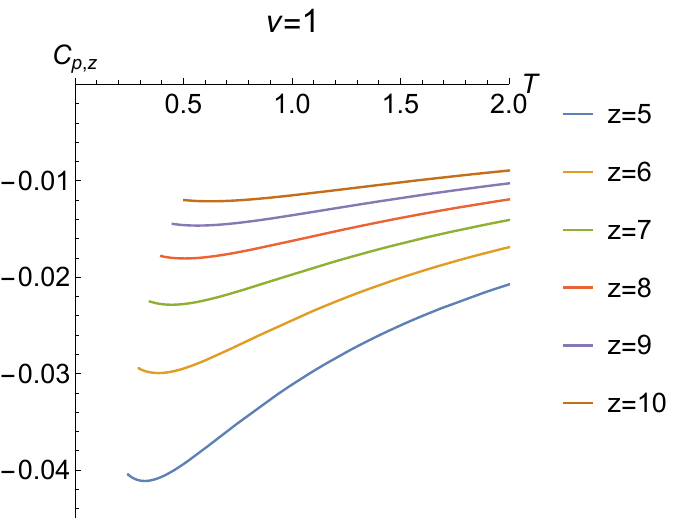}
         \caption{$C_{p,z}$ {\it vs.} $T$ for $\nu=1$ and various fixed $z$.}
         \label{fig:Cpplotfizedz4}
     \end{subfigure}
     \hfill
     \begin{subfigure}[h]{0.40\textwidth}
         \centering
         \includegraphics[width=\textwidth]{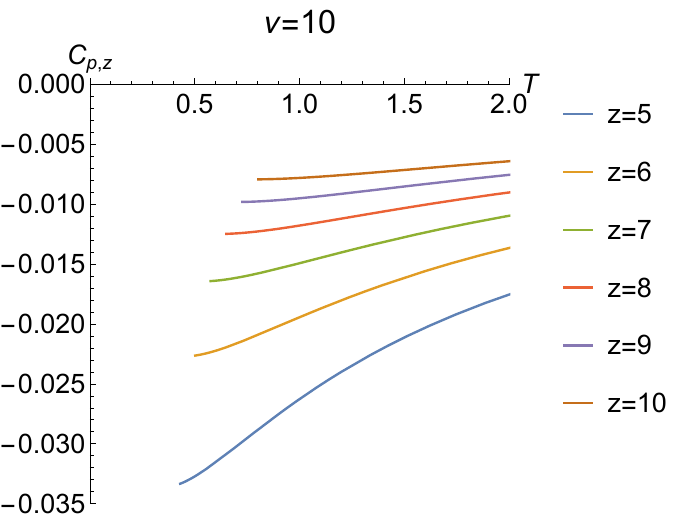}
         \caption{$C_{p,z}$ {\it vs.} $T$ for $\nu=10$ and various fixed $z$.}
         \label{fig:Cpplotfizedz5}
     \end{subfigure}
        \caption{$C_{p,z}$ {\it vs.} $T$ for various fixed values of $\nu$ and $z$. Note that figures \ref{fig:Cpplotfizedz2} and \ref{fig:Cpplotfizedz4} correspond to the positive and negative branches seen in figures \ref{fig:Cpplotfixedz1} and \ref{fig:Cpplotfixedz3}, respectively. Figure \ref{fig:Cpplotfizedz5} lacks positive branches because they are absent for those values of $\nu$ and $z$.}
        \label{fig:Cpplotfixedz}
\end{figure*}

 We provide a few $C_{p,z}$ plots in figure \ref{fig:Cpplotfixedz}.

As in the discussion on $C_{p,\alpha}$, it is difficult to make definitive general statements about $C_{p,z}$ due to the highly nontrivial functional form of the heat capacity. However, plotting the heat capacity as a function of temperature in this case reveals some interesting features. It can be seen in figure \ref{fig:Cpplotfixedz} that for sufficiently low values of $\nu$, the heat capacities contain a positive branch as well as a negative branch. Eventually, however, as $\nu$ rises sufficiently, the positive branch disappears and only the negative one remains. Thus, as the backreaction strength increases, the stable black holes are lost and only the unstable ones can exist. We again caution the reader that the regime of strong backreaction presents interpretational difficulties.

We comment here that in the small $\nu$ and small $z$ regime, the plots resemble straight, negatively-sloped lines in the third quadrant for $T\geq 0$. It is only after $z$ grows sufficiently that the different branches become apparent. For that reason, we chose to only plot the higher $z$ regime and omitted the low $z$ regime in our plots.

Of course, we made the easiest, simplest choice possible when calculating $C_p$ by assuming that either $z$ or $\alpha$ be held totally fixed, and only the other one be allowed to vary. An interesting question remains: how do different choices of $dz/d\lambda$ and $d\alpha/d\lambda$ in (\ref{Cpgeneral}) affect the functional behavior of $C_p(T)$? It is an interesting question we shall not tackle here, but the potential for generating many interesting and useful heat capacities lies in these choices.

\subsection{Specific heats at constant volume}
We now consider an analogous case to the one presented in the previous section. First we will consider for a moment the special case when $\alpha{=}0$, discussed at length in ref. \cite{Johnson:2023dtf}. Fixing $V$ to some constant value when $\alpha{=}0$ imposes a constraint between the independent coordinates $\nu$ and $z$. Since $V{=}V(\nu,z,\alpha{=}0)$ is a function of two independent variables, \emph{i.e.} a two-dimensional surface, fixing $V$ is geometrically equivalent to slicing this surface with a plane at constant $V$. Their intersection results in a one-dimensional isochoric curve, and one may evaluate $S$ and $T$ along the isochor. It is then trivial to calculate the appropriate derivative numerically pointwise. Because this calculation is being performed along a one-dimensional curve, there is never any ambiguity about directionality, as there is only one way to go on a one-dimensional curve (reversing direction and going the other way does not affect the value of the derivative, as both the change in $S$ and the change in $T$ pick up negative signs, which cancel). In the case of finite $\alpha$, the situation becomes more complex. Now, $V(\nu,z,\alpha)$ becomes a three-dimensional surface, and the constraint of constant $V$ necessitates intersecting this 3-surface with a 3-plane, whose intersection is a two-dimensional surface. This 2-surface is comprised of all the ordered triplets $(\nu,z,\alpha)$ such that they yield a fixed number when they are evaluated on $V$, and we may further recast them and consider them to correspond to, say, some relation of the form $\alpha{=}\alpha(\nu,z)$. Of course, we could have picked any of the the three variables to play the role of the dependent variable. The resulting surface may turn out to be multivalued, and this potential subtlety must be dealt with appropriately. Every single point on this surface, when evaluated on $V$, will yield the same constant number, so how does one pick a direction? It turns out that, just as in the constant pressure case, one is free to pick any direction one wishes, and different choices result in different $C_V$ functions.

We wish to calculate $C_V$ generally by combining $T$, $dS$, and $dT$ appropriately as before, but this time imposing the constraint that $dV{=}0$ as well. We will arrive at a more general formula by eschewing the coordinates $\nu$, $z$, and $\alpha$ temporarily and instead replacing them with $x^1$, $x^2$, and $x^3$, where the superscripts refer to indices, not exponents. We find that $C_V$ takes the form

\begin{widetext}
    \begin{align}
        \label{CVgeneral}
        C_V&(x^1,x^2,x^3)=T(x^1,x^2,x^3) \\ &\nonumber \times \left[ {\cal A}_\text{1}(x^1,x^2,x^3) \frac{dx^2}{d\lambda}+{\cal A}_\text{2}(x^1,x^2,x^3) \frac{dx^3}{d\lambda}\right] \left[{\cal B}_\text{1}(x^1,x^2,x^3) \frac{dx^2}{d\lambda}+{\cal B}_\text{2}(x^1,x^2,x^3) \frac{dx^3}{d\lambda}\right]^{-1},
        \end{align}
        
     \noindent where

     \begin{equation}
     \label{A1definition}
       {\cal A}_\text{1}(x^1,x^2,x^3)\equiv 
        \frac{\partial S}{\partial x^2}\bigg\rvert_{x^1,x^3}-\frac{\partial S}{\partial x^1}\bigg\rvert_{x^2,x^3} \left(\frac{\partial V}{\partial x^1}\bigg\rvert_{x^2,x^3}\right)^{-1} \frac{\partial V}{\partial x^2}\bigg\rvert_{x^1,x^3},
        \end{equation}

    \begin{equation}
    \label{A2definition}
       {\cal A}_\text{2}(x^1,x^2,x^3)\equiv \frac{\partial S}{\partial x^3}\bigg\rvert_{x^1,x^2}-\frac{\partial S}{\partial x^1}\bigg\rvert_{x^2,x^3} \left(\frac{\partial V}{\partial x^1}\bigg\rvert_{x^2,x^3}\right)^{-1} \frac{\partial V}{\partial x^3}\bigg\rvert_{x^1,x^2},
    \end{equation}

\begin{equation}
    \label{B1definition}
       {\cal B}_\text{1}(x^1,x^2,x^3)\equiv 
        \frac{\partial T}{\partial x^2}\bigg\rvert_{x^1,x^3}-\frac{\partial T}{\partial x^1}\bigg\rvert_{x^2,x^3} \left(\frac{\partial V}{\partial x^1}\bigg\rvert_{x^2,x^3}\right)^{-1} \frac{\partial V}{\partial x^2}\bigg\rvert_{x^1,x^3},
    \end{equation}

    \noindent and
\begin{equation}
    \label{B2definition}
       {\cal B}_\text{2}(x^1,x^2,x^3)\equiv \frac{\partial T}{\partial x^3}\bigg\rvert_{x^1,x^2}-\frac{\partial T}{\partial x^1}\bigg\rvert_{x^2,x^3} \left(\frac{\partial V}{\partial x^1}\bigg\rvert_{x^2,x^3}\right)^{-1} \frac{\partial V}{\partial x^3}\bigg\rvert_{x^1,x^2}.
    \end{equation}
    
\end{widetext}

The benefit in writing (\ref{CVgeneral})-(\ref{B2definition}) in this form is that we can now choose to assign each $x^i$ to correspond to any of the independent coordinates $\nu$, $z$, and $\alpha$, at our convenience. Regardless of the choice, the equation takes the same form and yields a valid heat capacity. We will see this borne out in the following subsections.

\begin{figure*}
     \centering
     \begin{subfigure}[h]{0.49\textwidth}
         \centering
         \includegraphics[width=\textwidth]{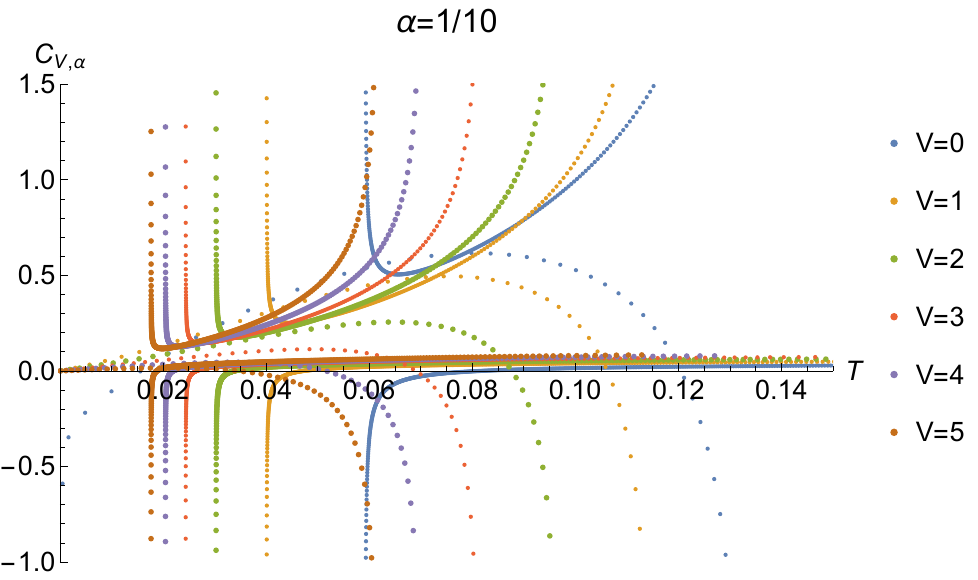}
         \caption{$C_{V,\alpha}$ {\it vs.} $T$ for $\alpha=1/10$ and various fixed $V$.}
         \label{fig:CValphaplot1}
     \end{subfigure}
     \begin{subfigure}[h]{0.49\textwidth}
         \centering
         \includegraphics[width=\textwidth]{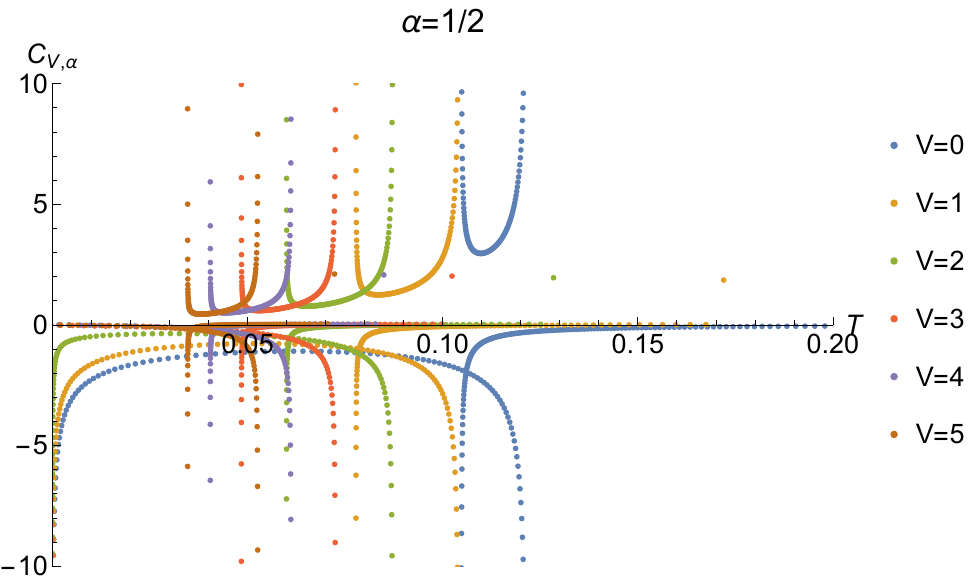}
         \caption{$C_{V,\alpha}$ {\it vs.} $T$ for $\alpha=1/2$ and various fixed $V$.}
         \label{fig:CValphaplot2}
         \end{subfigure}
     \hfill
     \begin{subfigure}[h]{0.49\textwidth}
         \centering
         \includegraphics[width=\textwidth]{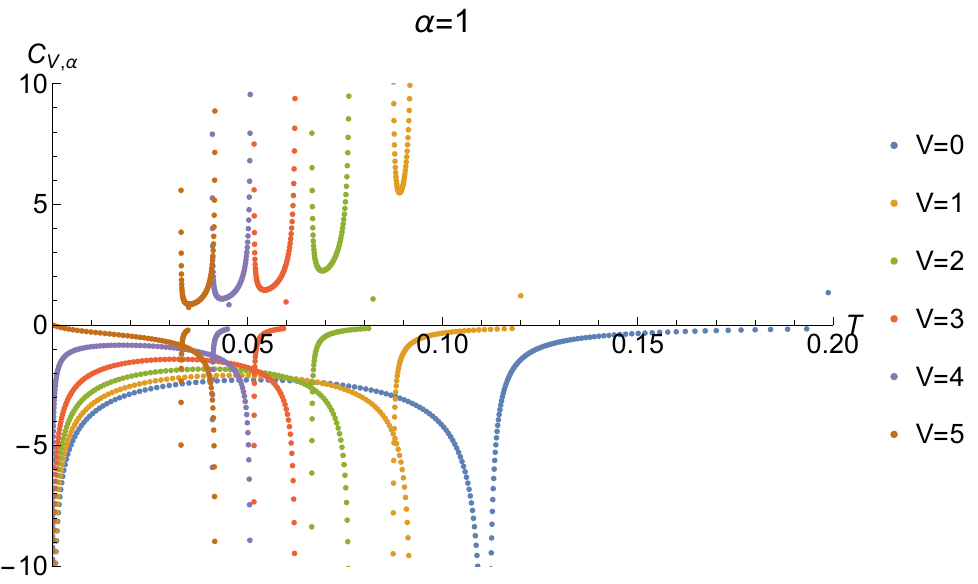}
         \caption{$C_{V,\alpha}$ {\it vs.} $T$ for $\alpha=1$ and various fixed $V$.}
         \label{fig:CValphaplot3}
     \end{subfigure}
        \caption{$C_{V,\alpha}$ {\it vs.} $T$ for various fixed values of $V$ and $\alpha$. Three branches are observed, with two temperatures at which divergences occur.}
        \label{fig:CVvsTfixedalpha}
\end{figure*}

However, before proceeding, it is crucial to underscore that to interpret (\ref{CVgeneral}) correctly, we must still restrict ourselves to values of $\nu$, $z$, and $\alpha$ which lie on an isochoric surface. Further restrictions allow us to constrain our consideration to a particular, albeit arbitrary, curve embedded within this isochoric surface, and $C_V$ may be calculated unambiguously along this curve. For the sake of simplicity, in this Paper we shall restrict ourselves to considering $C_V$ along curves of constant values of $\nu$, $z$, and $\alpha$, though we stress that these examples comprise only a tiny subset of all possible heat capacities at fixed $V$ the qBTZ can display. 

Further, when one of the independent variables is fixed, say $x^1$, (\ref{alphaconstraint}) defines a region in the $x^2{-}x^3$ plane, etc. Only the portions of isochors found within this region are physically realizable, and all others must be excluded, which sometimes entails truncating isochors when calculating $C_V$. All plotted heat capacities in the following subsections have only been plotted for isochors and portions of isochors which satisfy (\ref{alphaconstraint}), and we are thus assured that they are physical.

We note here that some analytic work is possible, though it is limited. For $V{=}0$, one can solve for the explicit function $\nu{=}\nu(z,\alpha)$, though its form is very complicated and not very illuminating, so we do not reproduce it here. The same can be done with $\alpha{=}\alpha(\nu,z)$. In both cases, four solutions exist, and one must pick the appropriate one. These functions may be composed with the appropriate $C_{V}$ function to generate some of the plots generated numerically in the figures contained in the following sections. In some cases the heat capacities required stitching two or more of these solutions together to generate the full curve, which contributed to our decision to generate the curves fully numerically instead.

However, we did explore the analytical route, and we make here some relevant observations. Some of the most interesting features of the heat capacities is the presence of potential divergences which may signal some type of phase transition, and we expect the derivative of $T$ to be zero there. One may compose the above functions with $T$ to plot the temperature at $V{=}0$, and we indeed noted the presence of points where the temperature function has zero slope, corresponding to these divergent points. We sought exact results of interesting points in the $C_V$ plots using this avenue, and while the points satisfying these criteria were observed, the complexity of the equations made it very difficult to find their values exactly.

\subsubsection{$C_V$ at fixed $\alpha$}

We begin with the case of fixed $\alpha$, as it is the case most closely related to our previous work in ref. \cite{Johnson:2023dtf}, where $\alpha{=}0$. In that case, for different values of fixed $V$, $C_V$ contained two branches. The first branch originated at $T{=}0$ and was initially positive, before reaching a maximum and then becoming negative and diverging to negative infinity at some temperature $T_2$. The second branch originated at at some temperature $T_1$, where $0<T_1<T_2$, rather than at absolute zero, and it remained positive always. It too diverged at $T_2$ as the first branch, except it diverged to positive infinity. To investigate this more general case, we will set $x^1{=}\nu$, $x^2{=}z$, and $x^3{=}\alpha$. The heat capacity is then given by
\begin{widetext}
\begin{align}
    \label{CVfixedalpha}
    C_{V,\alpha}&=T(\nu,z,\alpha)\\\times&\nonumber \left[ \frac{\partial S}{\partial z}\bigg\rvert_{\nu,\alpha}-\frac{\partial S}{\partial \nu}\bigg\rvert_{z,\alpha} \left(\frac{\partial V}{\partial \nu}\bigg\rvert_{z,\alpha}\right)^{-1} \frac{\partial V}{\partial z}\bigg\rvert_{\nu,\alpha}\right] \left[ \frac{\partial T}{\partial z}\bigg\rvert_{\nu,\alpha}-\frac{\partial T}{\partial \nu}\bigg\rvert_{z,\alpha} \left(\frac{\partial V}{\partial \nu}\bigg\rvert_{z,\alpha}\right)^{-1} \frac{\partial V}{\partial z}\bigg\rvert_{\nu,\alpha}\right]^{-1}.
\end{align}
\end{widetext}

\noindent The explicit form of (\ref{CVfixedalpha}) may be found in Appendix \ref{appendixexplicitquantities}.

For finite $\alpha$, we see three branches in the heat capacity. The first two seemingly correspond to the branches seen in the $\alpha{=}0$ case, with some differences. The first branch resembles one of the branches seen in the $\alpha{=}0$ most. It is zero at $T{=}0$ for most volumes at low $\alpha$, except for $V{=}0$, in which case it diverges to negative infinity as $T{\rightarrow} 0^{+}$. As $T$ rises, the first branch of $C_{V,\alpha}$ reaches a maximum and then begins decreasing and becomes negative, diverging to negative infinity as it approaches some maximum $T_2$, not necessarily the same value as in the $\alpha{=}0$ case. The second branch resembles the other branch seen in the $\alpha{=}0$ case, also diverging at $T_2$, except to positive infinity. However, rather than decreasing to some minimum $T_1$ and simply terminating, as in the static case, its other end also diverges to positive infinity at some minimum value $T_1$, where $0{<}T_1{<}T_2$. The third branch, which has no static analog, is negative for low values of $T$ and also diverges to negative infinity as $T{\rightarrow} T_1^{+}$. As $T$ rises, the third branch becomes less negative, and for some values of $\alpha$ it becomes positive. No further divergences are observed in the range of $T$ numerically explored in figure \ref{fig:CVvsTfixedalpha}. Note that as $\alpha$ is raised, the first branch drops to lower values, eventually becoming fully negative, while its behavior at $T{=}0$ also resembles the $V{=}0$ case for low $\alpha$, where a divergence arises as $T{\rightarrow} 0^{+}$. The second branch rises in value and its $T$ range becomes smaller as $\alpha$ increases, or put differently, $T_1$ and $T_2$ approach each other. The relative divergent behavior of the branches is unchanged, with both the first and second branches diverging to negative infinity and positive infinity, respectively, at $T_2$, and the lower end of the second and third branches diverging to positive and negative infinity, respectively, at $T_1$. Note that for the two greater values of $\alpha$ we considered in figure \ref{fig:CVvsTfixedalpha}, the third branch does not become positive for the values of $T$ shown in the plots.

For fixed $\alpha$ and $V{=}0$, where some limited analytical work is possible, as mentioned above, one can clearly see that the temperature function resembles the $\alpha{=}0$ case, which is analyzed in depth in refs. \cite{Johnson:2023dtf,Frassino:2023wpc}. Specifically, as $z$ rises, the temperature rises to a local maximum, then drops to a local minimum, before rising again. These two local extrema correspond to the two divergences noted in the $C_{V,\alpha}$ plots. The exact values of $z$ where these extrema are found are very difficult to calculate analytically, and thus the temperatures at which the heat capacity diverges is hard to find, but as $\alpha$ is varied it becomes visually apparent that these divergences are a feature present over many values of $\alpha$.

\begin{figure*}
     \centering
     \begin{subfigure}[h]{0.49\textwidth}
         \centering
         \includegraphics[width=\textwidth]{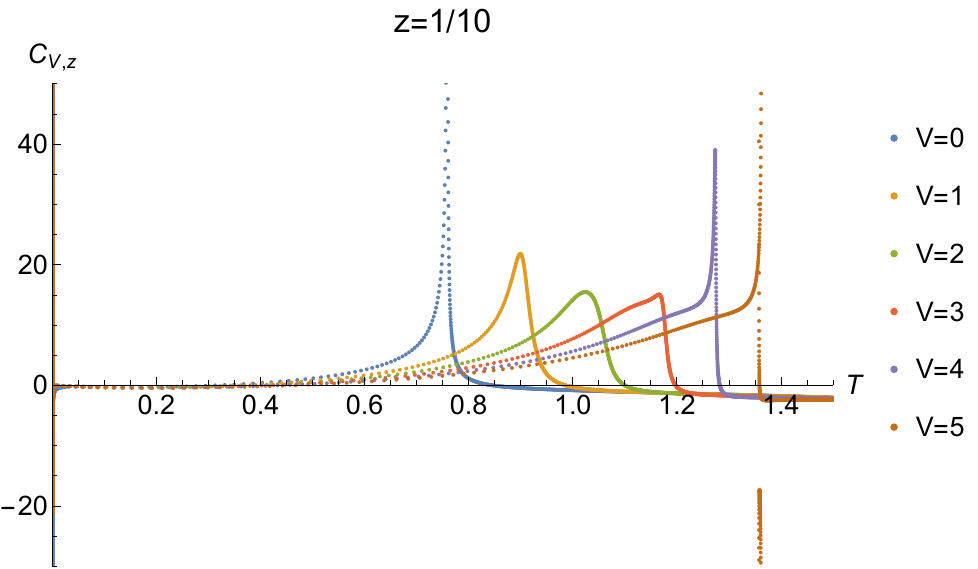}
         \caption{$C_{V,z}$ {\it vs.} $T$ for $z{=}1/10$ and various fixed $V$.}
         \label{fig:CVzplot1}
     \end{subfigure}
     \hfill
     \begin{subfigure}[h]{0.49\textwidth}
         \centering
         \includegraphics[width=\textwidth]{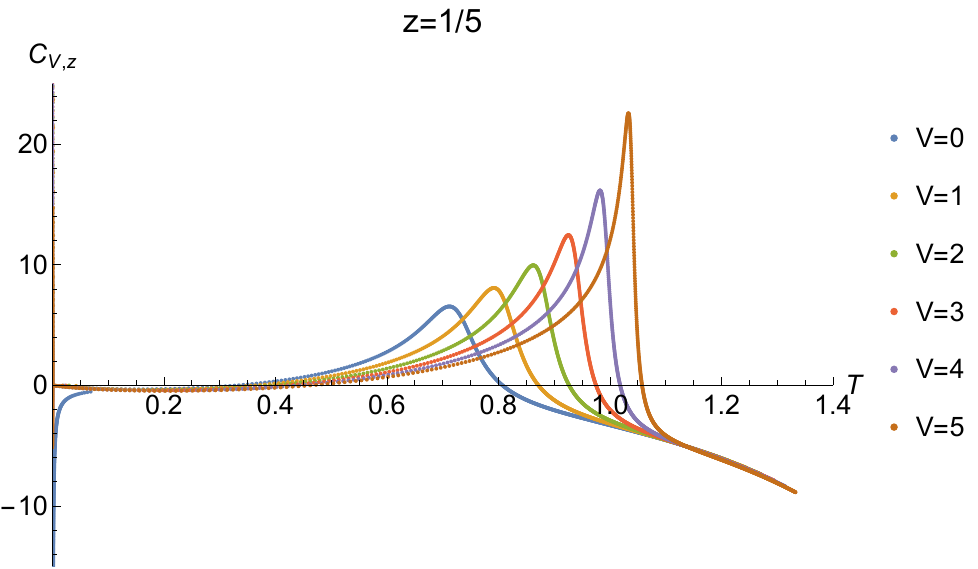}
         \caption{$C_{V,z}$ {\it vs.} $T$ for $z=1/5$ and various fixed $V$.}
         \label{fig:CVzplot2}
         \end{subfigure}
     \hfill
     \begin{subfigure}[h]{0.49\textwidth}
         \centering
         \includegraphics[width=\textwidth]{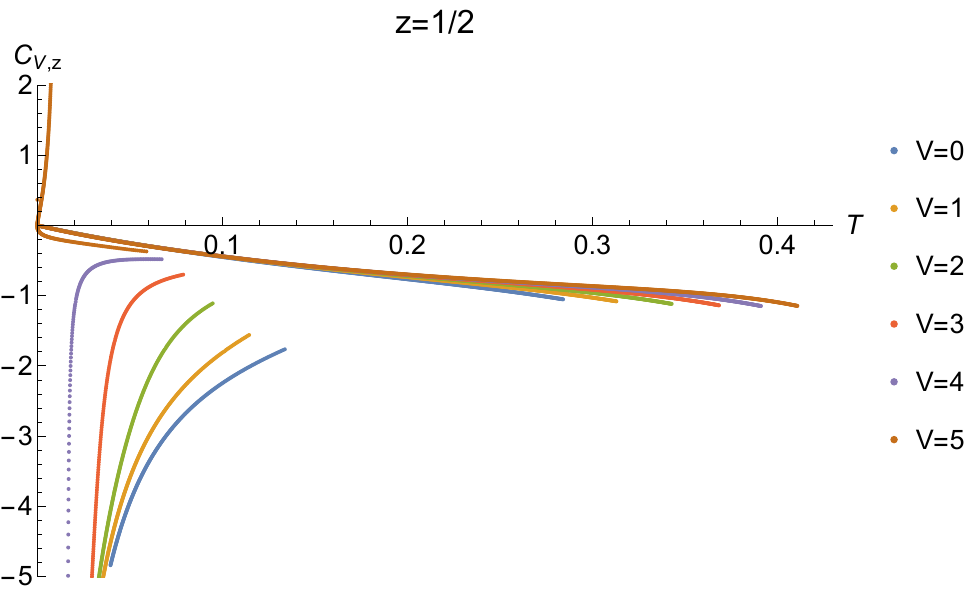}
         \caption{$C_{V,z}$ {\it vs.} $T$ for $z{=}1/2$ and various fixed $V$.}
         \label{fig:CVzplot3}
         \end{subfigure}
         \hfill
     \begin{subfigure}[h]{0.49\textwidth}
         \centering
         \includegraphics[width=\textwidth]{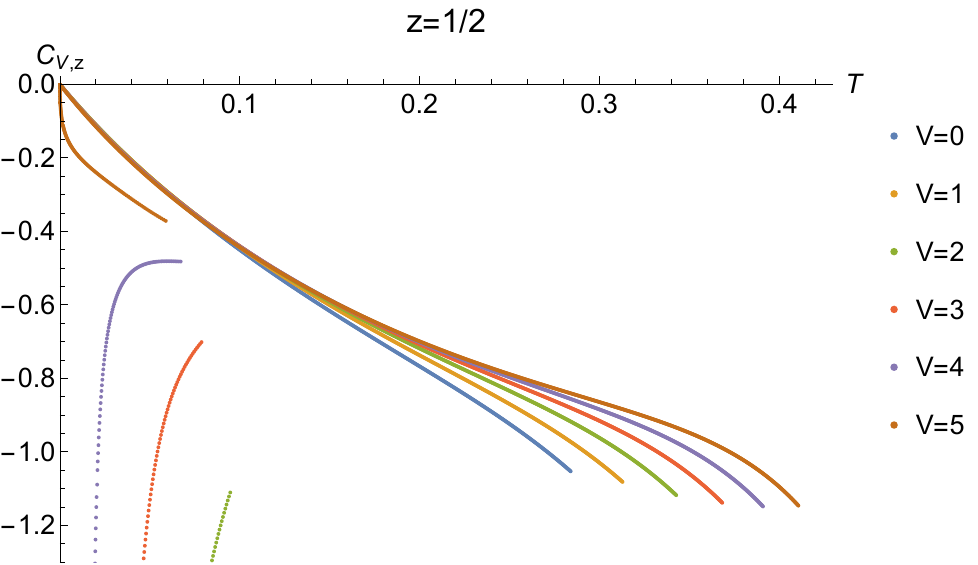}
         \caption{$C_{V,z}$ {\it vs.} $T$ for $z{=}1/2$ and various fixed $V$.}
         \label{fig:CVzplot3two}
     \end{subfigure}
        \caption{$C_{V,z}$ {\it vs.} $T$ for various fixed values of $V$ and $z$. Note that figure \ref{fig:CVzplot3two} corresponds to the upper negative branches seen in figure \ref{fig:CVzplot3}.}
        \label{fig:CVvsTfixedz}
\end{figure*}

\subsubsection{$C_V$ at fixed $z$}

We may also calculate $C_V$ in the case where $z$ is held fixed. We again make the choice $x^1{=}\nu$, $x^2{=}z$, and $x^3{=}\alpha$, and the resulting heat capacity is

\begin{widetext}
    \begin{align}
        \label{CVfixedz}        C_{V,z}&=T(\nu,z,\alpha)\\\times&\nonumber \left[ \frac{\partial S}{\partial \alpha}\bigg\rvert_{\nu,z}-\frac{\partial S}{\partial \nu}\bigg\rvert_{z,\alpha} \left(\frac{\partial V}{\partial \nu}\bigg\rvert_{z,\alpha}\right)^{-1} \frac{\partial V}{\partial \alpha}\bigg\rvert_{\nu,z}\right] \left[ \frac{\partial T}{\partial \alpha}\bigg\rvert_{\nu,z}-\frac{\partial T}{\partial \nu}\bigg\rvert_{z,\alpha} \left(\frac{\partial V}{\partial \nu}\bigg\rvert_{z,\alpha}\right)^{-1} \frac{\partial V}{\partial \alpha}\bigg\rvert_{\nu,z}\right]^{-1}.
    \end{align}
\end{widetext}

\noindent The explicit form of (\ref{CVfixedalpha}) may be found in Appendix \ref{appendixexplicitquantities}.

At fixed $z$, $C_V$ displays two physically realizable branches, one which is fully negative, and another one which contains both positive and negative regions as well as a local maximum which for certain fixed values of $z$ and $V$ approaches a divergence. The fully negative branch is only found at very small values of $T$, and may not be readily apparent in the plots for the two lower values of $z$ in figure \ref{fig:CVvsTfixedz}, though it becomes more pronounced at higher $z$. Note in figure \ref{fig:CVzplot1} that for $V{=}0$ and $z{=}1/10$ a divergence is forming, though numerical analysis of the temperature and its derivative show that while the slope of $T$ is small, it not zero here. As $V$ rises, the heat capacity again seems to be in the neighborhood of a divergence of sorts, though it is also apparent that multivaluedness is forming in the plots. A careful inspection reveals that the $S$ \emph{vs.} $T$ plot is also multivalued. At the higher value of $z{=}1/5$, these divergent and multivalued points are absent in the plots, and as $V$ grows we see the peaks becoming more pronounced, possibly signaling the formation of a divergence. For sufficiently high values of $z$, we see that the branches become fully negative, with the exception of $V{=}5$, which still possesses a positive branch at small $T$. The fully negative branches remain fully negative, but are now found at higher temperatures than before. AS before, a full analytic analysis of these interesting divergent points in the heat capacity are very difficult to perform exactly due to the highly complex nature of the heat capacity.

\subsubsection{$C_V$ at fixed $p$}

\begin{figure*}
     \centering
     \begin{subfigure}[h]{0.49\textwidth}
         \centering
         \includegraphics[width=\textwidth]{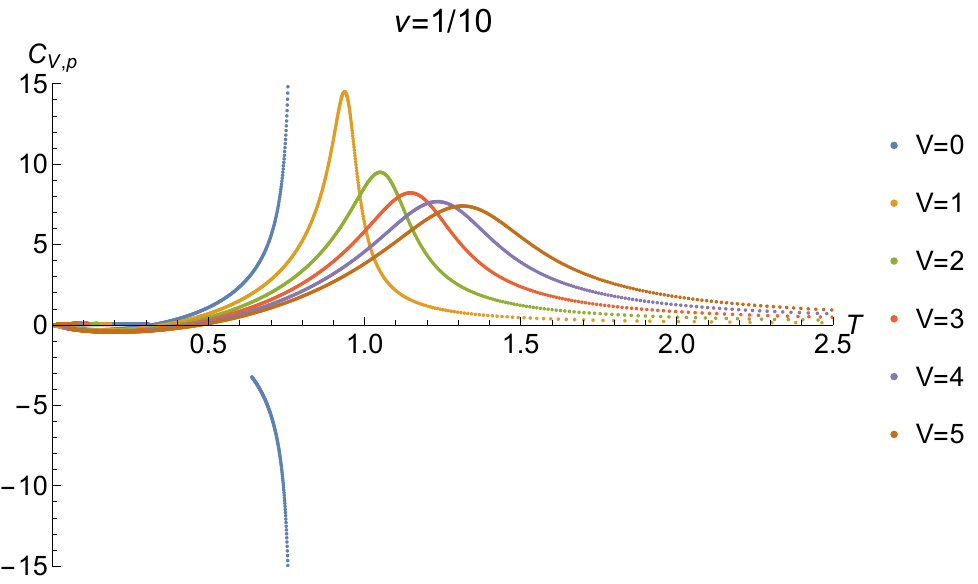}
         \caption{$C_{V,p}$ {\it vs.} $T$ for $\nu{=}1/10$ and various fixed $V$.}
         \label{fig:CVpplot1}
     \end{subfigure}
     \hfill
     \begin{subfigure}[h]{0.49\textwidth}
         \centering
         \includegraphics[width=\textwidth]{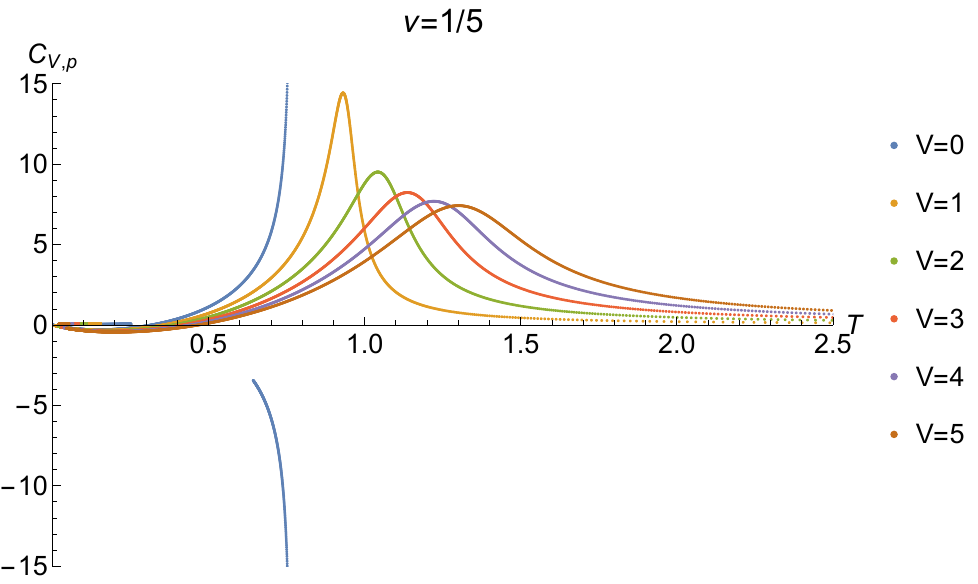}
         \caption{$C_{V,p}$ {\it vs.} $T$ for $\nu{=}1/5$ and various fixed $V$.}
         \label{fig:CVpplot2}
         \end{subfigure}
     \hfill
     \begin{subfigure}[h]{0.49\textwidth}
         \centering
         \includegraphics[width=\textwidth]{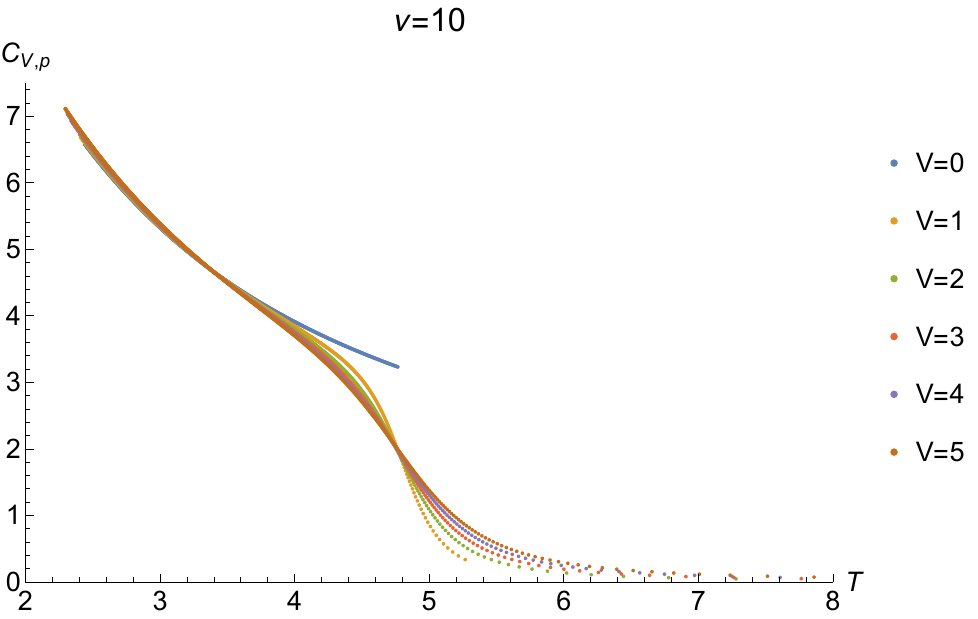}
         \caption{$C_{V,p}$ {\it vs.} $T$ for $\nu{=}10$ and various fixed $V$.}
         \label{fig:CVpplot3}
         \end{subfigure}
         \hfill
     \begin{subfigure}[h]{0.49\textwidth}
         \centering
         \includegraphics[width=\textwidth]{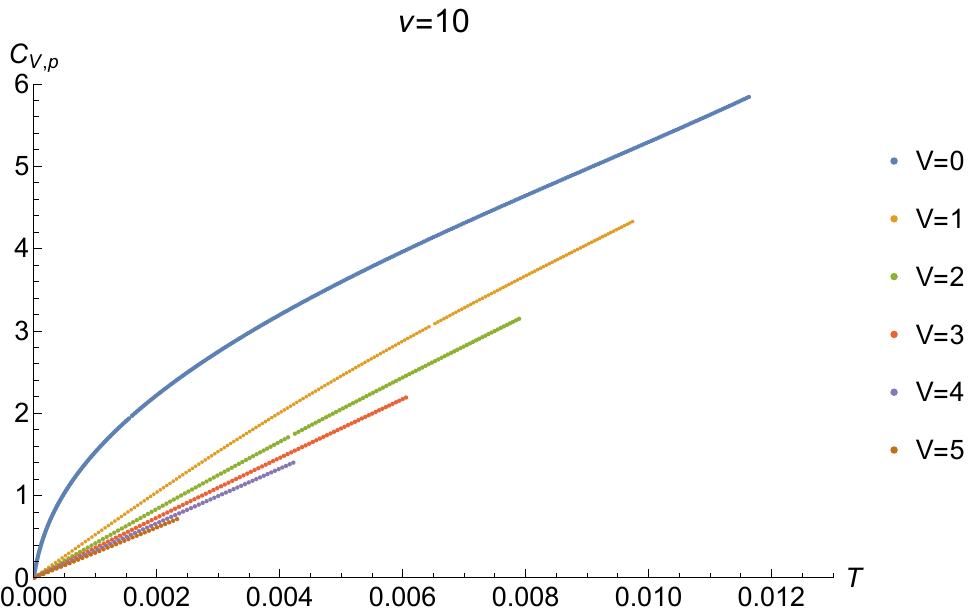}
         \caption{$C_{V,p}$ {\it vs.} $T$ for $\nu{=}10$ and various fixed $V$.}
         \label{fig:CVpplot4}
     \end{subfigure}
        \caption{$C_{V,p}$ {\it vs.} $T$ for various fixed values of $V$ and $p$. Note that figures \ref{fig:CVpplot3} and \ref{fig:CVpplot4} correspond to the same $p$, though the latter plots branches seen only at very low $T$, not plotted in the former.}
        \label{fig:CVvsTfixedp}
\end{figure*}

Finally, we present $C_V$ when $p$, and thus $\nu$, is held fixed. This time\footnote{Readers may have noticed that a different choice of the $x^i$ will result in a heat capacity with the derivatives swapped around compared to (\ref{CVfixedp}). This is true, but the results are nevertheless all equivalent to one another when explicitly evaluated.}, we make the choice $x^1{=}\alpha$, $x^2{=}z$, $x^3{=}\nu$. The resulting heat capacity is given by

\begin{widetext}
    \begin{align}
        \label{CVfixedp}        C_{V,p}&=T(\nu,z,\alpha)\\\times&\nonumber \left[ \frac{\partial S}{\partial z}\bigg\rvert_{\nu,\alpha}-\frac{\partial S}{\partial \alpha}\bigg\rvert_{\nu,z} \left(\frac{\partial V}{\partial \alpha}\bigg\rvert_{\nu,z}\right)^{-1} \frac{\partial V}{\partial z}\bigg\rvert_{\nu,\alpha}\right] \left[ \frac{\partial T}{\partial z}\bigg\rvert_{\nu,\alpha}-\frac{\partial T}{\partial \alpha}\bigg\rvert_{\nu,z} \left(\frac{\partial V}{\partial \alpha}\bigg\rvert_{\nu,z}\right)^{-1} \frac{\partial V}{\partial z}\bigg\rvert_{\nu,\alpha}\right]^{-1}.
    \end{align}
\end{widetext}

\noindent The explicit form of (\ref{CVfixedp}) may be found in Appendix \ref{appendixexplicitquantities}.

For low values of $\nu$ investigated, the heat capacity turned out to consist of two branches, which we will refer to as ``small" and ``large," with the small branch having a small magnitude compared to the large branch. The exception is the case $V{=}0$, where three branches can be seen, one small and two large. Of the two large branches, one is positive and the other is negative. In the two lower values of $\nu$ plotted, the small branch was found at very low temperatures only, though they are not readily apparent in figures \ref{fig:CVpplot1} and \ref{fig:CVpplot2} due to their small magnitude compared to the large branches. The small branches are also found at the higher value of $\nu$ considered, and while they are still restricted to very low temperatures, they are no longer small in comparison to the so-called large branches. We provide a close-up plot of these ``small" branches at higher $\nu$ in figure \ref{fig:CVpplot4}. For values of the volume above zero in the large branches, very similarly to the fixed $z$ case, we observe in the heat capacities what appears to be the formation of a sharpening maximum which has the potential to tend to a divergence in the neighborhood of $V{=}0$ and $\nu{=}1/10$, and as $V$ rises, the curves flatten out. As usual, the functional complexity of the heat capacities makes exact, analytic results very difficult to come by.

\section{Super entropicity and the signs of the heat capacities}

We return now to the question of whether there is a cogent relationship between the superentropicity of the qBTZ black hole and its stability vis-\`a-vis the signs of its heat capacities. Of course, this question may be investigated to all orders in $\nu$, and doing so revealed no obvious relationship between the value of $\mathcal R$ and the signs of the heat capacities considered above. However, we must again emphasize that it is unclear whether the relevant equations may be meaningfully wielded to all orders in the backreaction, and it is perhaps more instructive to remain in the limit of small $\nu$, where the theory of gravity induced on the brane is nearly massless and the applicable rules are well understood.

Further, the presence of a very large family of heat capacities at either fixed $p$ or $V$ in the rotating case complicates things somewhat. Unlike in the static case, where there was a single $C_p$ or $C_V$, we now find ourselves with if not an infinite number of each, at least a very large quantity far exceeding the few special cases considered in this Paper. As was underscored above, every new path chosen in the phase space of allowed states corresponds to a new heat capacity. Each of these would have to be analyzed in order to completely address this question, though there is no obvious straightforward way to do this in general for arbitrary paths in the phase space.

However, based solely on the heat capacities discussed in this Paper, it is still be possible to reach some simple conclusions. To second order in $\nu$,

\begin{equation}
    \label{Rsmallnu}
    {\cal R}\simeq 1+f(z,\alpha) \nu-g(z,\alpha)\nu^2+O(\nu^3).
\end{equation}

\noindent where

\begin{equation}
    \label{Rf}
    f(z,\alpha)=\frac{\alpha ^2+2 \alpha ^2 z^4-\alpha ^2 z^2-2 z^2}{2 z \left(-\alpha ^2+\alpha ^2 z^2-1\right)}
\end{equation}

\noindent and
\begin{widetext}
    \begin{equation}
        \label{Rg}
        g(z,\alpha)=\frac{\alpha ^4-2 \alpha ^2+8 \alpha ^4 z^8+8 \alpha ^4 z^6-12 \alpha ^2 z^6-7 \alpha ^4 z^4-18 \alpha ^2 z^4+4 z^4+2 \alpha ^4 z^2+6 z^2}{8 z^2
   \left(-\alpha ^2+\alpha ^2 z^2-1\right)^2}.
    \end{equation}
\end{widetext}

Due to the complexity of $f$ and $g$, expanded in small $\nu$, even only considering $f$, it is possible to find values of the parameters for which all the heat capacities are negative, while nevertheless $\mathcal R{>}\text{1}$. All other combination of relative heat capacity signs and values for $\mathcal{R}$ can also be found if one hunts around for them, leading us to the conclusion that in this system there is no apparent relationship between the superentropicity and the stability of the black hole. If such a relationship exists, it must not be as straightforward as conjectured.

 \begin{acknowledgments}
The auhor wishes to acknowledge CVJ for his help and support during the writing of this manuscript.   
\end{acknowledgments}

\appendix

\section{Calculating $M$, $T$, $S$, $J$, and $\Omega$}
\label{appendixMTSJO}

We provide here abridged calculations for $M$, $T$, $S$, $J$, and $\Omega$. The full details of the calculations and discussions on the subtleties which arise are found in ref. \cite{Emparan:2020znc}. Recall that the setting described in the Paper is an AdS$_4$ C-metric with an AdS$_3$ brane at the location of the regulator surface. The AdS$_4$ C-metric takes the form
\begin{widetext}
\begin{equation}
    \label{C-metric}
    ds^2=\frac{\ell^2}{(\ell+xr)^2}\left[-\frac{H(r)}{\Sigma (x,r)}(dt+a x^2 d\phi)^2+\frac{\Sigma (x,r)}{H(r)}dr^2+r^2\left(\frac{\Sigma (x,r)}{G(x)}dx^2+\frac{G(x)}{\Sigma (x,r)}\left(d\phi-\frac{a}{r^2}dt\right) ^2\right)\right],
\end{equation}
\end{widetext}

\noindent where

\begin{equation}
    \label{Hfxn}
    H(r)=\frac{r^2}{\ell_3^2}+\varkappa-\frac{\mu\ell}{r}+\frac{a^2}{r^2},
\end{equation}

\begin{equation}
    \label{Gfxn}
    G(x)=1-\varkappa x^2-\mu x^3+\frac{a^2}{\ell_3^2}x^4,
\end{equation}

\noindent and

\begin{equation}
    \label{Sigmafxn}
    \Sigma(x,r)=1+\frac{a^2x^2}{r^2}.
\end{equation}

\noindent Note that in the preceding equations, $\mu{=}0$ corresponds to pure AdS and $\mu{>}0$ accounts for the holographic corrections to the quantum black hole.

The metric induced on the brane by the backreaction (the brane is fixed at $x{=}0$) naively (see below) takes the form

\begin{equation}
    \label{induced metric}
    ds^2{=}-H(r)dt^2+H(r)^{-1}dr^2+r^2 \left(d\phi-\frac{a}{r^2}dt\right)^2.
\end{equation}

 \noindent Note that in (\ref{induced metric}) a $2\pi\Delta$ periodicity is imposed upon $\phi$ to avoid a conical singularity, where

\begin{equation}
    \label{deltaterm}
    \Delta\equiv\frac{2x_1}{3-\varkappa x_1^2-\tilde{a}^2},
\end{equation}

\noindent and where we've defined the dimensionless parameter $\tilde{a}{\equiv}ax_1^2/\ell_3$. We call (\ref{induced metric}) the ``naive" form of the metric because because if one goes from $\phi$ to $\phi+2\pi\Delta$, one does not form a closed curve, since one ends up on a different constant time slice of spacetime than one started on. See ref. \cite{Emparan:2020znc} for details.

The coordinate transformation that eliminates this issue and makes the (new) angular coordinate $2\pi$ periodic is

\begin{eqnarray}
    \label{transformation}
    t&=&\Delta(\bar{t}-\tilde{a}\ell_3\bar{\phi}),\\
    \phi&=&\Delta\left(\bar{\phi}-\frac{\tilde{a}}{\ell_3}\bar{t}\right),\\
    r^2&=&\frac{\bar{r}^2-r_s^2}{(1-\tilde{a}^2)\Delta^2},
\end{eqnarray}

\noindent where

\begin{equation}
    \label{r_s}
    r_s\equiv\frac{\ell_3\tilde{a}\Delta}{x_1} \sqrt{2-\varkappa x_1^2}=\ell_3\frac{2\tilde{a}\sqrt{2-\varkappa x_1^2}}{3-\varkappa x_1^2-\tilde{a}^2}.
\end{equation}

We find $M$ and $J$ from the metric expressed in the new coordinates. The results are

\begin{eqnarray}
    \label{mass}
    M&=&{-}\frac{\varkappa\Delta^2}{8{\cal{G}}_3}\left(1+\tilde{a}^2-\frac{4\tilde{a}^2}{\varkappa x_1^2}\right)\\&=&\frac{1}{2{\cal{G}}_3}\frac{-\varkappa x_1^2+\tilde{a}^2(4-\varkappa x_1^2)}{(3-\varkappa x_1^2-\tilde{a}^2)^2}\nonumber
\end{eqnarray}

\noindent and

\begin{eqnarray}
    \label{angularmomentum}
    J&=&\frac{\ell_3}{4{\cal{G}}_3}\tilde{a}\mu x_1\Delta^2\\
    &=&\frac{\ell_3}{{\cal{G}}_3}\frac{\tilde{a}(1-\varkappa x_1^2+\tilde{a}^2)}{(3-\varkappa x_1^2-\tilde{a}^2)^2}, \nonumber
\end{eqnarray}

\noindent where we have employed the relation found in {\cite{Emparan:2020znc}} given by

\begin{equation}
    \label{mux1relation}
    \mu x_1=-\varkappa\frac{(1+z^2)(1+\alpha^2(1-z^2))}{1-\nu z^3}
\end{equation}

\noindent in (\ref{angularmomentum}). Further, using the result from {\cite{Emparan:2020znc}} that

\begin{eqnarray}
    \label{relations}
    -\varkappa x_1^2=\frac{1-\nu z^3}{z^2(1+\nu z-\alpha^2 z(z-\nu))},
\end{eqnarray}

\noindent the results {(\ref{qM})} and (\ref{qJ}) follow directly.

To find $S$, we simply use the Bekenstein-Hawking area law in the bulk, which gives us the quantum-corrected, generalized entropy, which also equals the standard entropy in this case, and find the area of the black hole using the modified metric after the coordinate transformations are applied \cite{Emparan:2020znc,Frassino:2022zaz,Emparan:2006ni}, then rewrite everything in terms of $D{=}3$ quantities:
\begin{widetext}
\begin{equation}
    \label{entropy}
    S=\frac{1}{4G_3}\int_{0}^{2\pi}d\bar{\phi}\int_{0}^{x_1} dx\frac{r_{+}^2\ell^2\Delta}{(\ell+r_{+}x)^2}\left(1+\frac{a^2x_1^2}{r_{+}^2}\right).
\end{equation}
\end{widetext}

\noindent Explicitly evaluating (\ref{entropy}) yields (\ref{qS}).

We will now depart from ref. \cite{Emparan:2020znc} in calculating the rest of the thermodynamic quantities. While the above reference contains a very interesting discussion on Killing vectors and how the isometries of the metric allow one to calculate these quantities, the results must be translated again into functions of $\nu$, $z$, and $\alpha$ to make use of them. We wish to derive these quantities directly from the thermodynamic quantities we already possess, since they yield some interesting results. Recall from the First Law that

\begin{equation}
\label{omeganiceform}
    \Omega=\frac{\partial M}{\partial J}\bigg\rvert_{p,S}.
\end{equation}

\noindent Constructing (\ref{omeganiceform}) from the variations of $M$ and $J$ while holding $p$ and $S$ fixed (see the appendix of \cite{Johnson:2023dtf} for details on this procedure) results in
\begin{widetext}
\begin{equation}
    \label{omegaformula}
    \Omega=\left[\frac{\partial M}{\partial \alpha}\bigg\rvert_{\nu,z}-\frac{\partial M}{\partial z}\bigg\rvert_{\nu,\alpha}\left(\frac{\partial S}{\partial z}\bigg\rvert_{\nu,\alpha}\right)^{-1}\frac{\partial S}{\partial \alpha}\bigg\rvert_{\nu,z}\right]\left[\frac{\partial J}{\partial \alpha}\bigg\rvert_{\nu,z}-\frac{\partial J}{\partial z}\bigg\rvert_{\nu,\alpha}\left(\frac{\partial S}{\partial z}\bigg\rvert_{\nu,\alpha}\right)^{-1}\frac{\partial S}{\partial \alpha}\bigg\rvert_{\nu,z}\right]^{-1}.
\end{equation}
\end{widetext}

\noindent Explicitly evaluating (\ref{omegaformula}) yields (\ref{qOmega}).

Similarly, we may calculate the temperature from the First Law as

\begin{equation}
    \label{Tniceform}
    T=\frac{\partial M}{\partial S}\bigg\rvert_{p,J}.
\end{equation}

\noindent Constructing (\ref{Tniceform}) from the variations of $M$ and $S$ while holding $p$ and $J$ fixed results in

\begin{widetext}
    \begin{equation}
        \label{Tformula}
        T=\left[\frac{\partial M}{\partial \alpha}\bigg\rvert_{\nu,z}-\frac{\partial M}{\partial z}\bigg\rvert_{\nu,\alpha}\left(\frac{\partial J}{\partial z}\bigg\rvert_{\nu,\alpha}\right)^{-1}\frac{\partial J}{\partial \alpha}\bigg\rvert_{\nu,z}\right]\left[\frac{\partial S}{\partial \alpha}\bigg\rvert_{\nu,z}-\frac{\partial S}{\partial z}\bigg\rvert_{\nu,\alpha}\left(\frac{\partial J}{\partial z}\bigg\rvert_{\nu,\alpha}\right)^{-1}\frac{\partial J}{\partial \alpha}\bigg\rvert_{\nu,z}\right]^{-1}.
    \end{equation}
\end{widetext}

\noindent Explicitly evaluating (\ref{Tformula}) yields (\ref{qT}).
\begin{widetext}
\section{Some quantities calculated explicitly}
\label{appendixexplicitquantities}

We present in this appendix explicit forms of the functions $V$, $C_{p,\alpha}$, $C_{p,z}$, $C_{V,\alpha}$, $C_{V,z}$, and $C_{V,p}$  as functions of the three dimensionless parameters of the theory: $\nu$, $z$, and $\alpha$. We begin with $V$, given by

\begin{equation}
    V(\nu,z,\alpha)=\frac{{\cal N}_\text{1}(\nu, z,\alpha)}{{\cal D}_\text{1}(\nu, z,\alpha)},
\end{equation}

\noindent where
  
    \begin{equation}
    \begin{split}
    \label{volumenumerator}
        {\cal {N}}_1 (\nu,z,\alpha)\equiv &-2\pi \ell_3^2(\nu) \left(-\alpha ^2 \nu ^2+2 \alpha ^4 \nu ^2 z^8-6 \alpha ^4 \nu ^3 z^7-4 \alpha ^4 \nu  z^7-3 \alpha ^2 \nu ^3 z^7+4 \alpha ^4 \nu ^4 z^6+4 \alpha ^4 \nu ^2 z^6\right.\\&-2 \alpha ^4 z^6+4 \alpha ^2 \nu ^4 z^6-2 \alpha ^2 \nu ^2 z^6+\nu ^4 z^6+6 \alpha ^4 \nu  z^5+6 \alpha ^2 \nu ^3 z^5+8 \alpha ^2 \nu  z^5+3 \nu ^3 z^5-2 \alpha ^4 \nu ^2 z^4\\&+4 \alpha ^4 z^4-7 \alpha ^2 \nu ^2 z^4+4 \alpha ^2 z^4-2 \alpha ^4 \nu ^3 z^3-4 \alpha ^4 \nu  z^3+\alpha ^2 \nu ^3 z^3-6 \alpha ^2 \nu  z^3+\nu ^3 z^3-4 \nu  z^3\\&\left.-2 \alpha ^4 z^2-2 \alpha ^2 \nu ^2 z^2-4 \alpha ^2 z^2+\nu ^2 z^2-2 z^2+2 \alpha ^4 \nu  z+2 \alpha ^2 \nu  z \right)
    \end{split}
    \end{equation}
  
\noindent and

\begin{equation}
    \label{volumedenominator}
    {\cal D}_\text{1}(\nu,z,\alpha)\equiv \left(1+3z^2+2\nu z^3+\alpha^2\left(4\nu z^3-1-3z^4\right)\right)^2.
\end{equation}

$C_{p,\alpha}$ is given by

\begin{equation}
     C_{p,\alpha}(\nu,z,\alpha)=\frac{{\cal N}_\text{2}(\nu, z,\alpha)}{{\cal D}_\text{2}(\nu, z,\alpha)},
\end{equation}

\noindent where
\begin{equation}
\begin{split}
    \label{Cpalphanumerator}
    {\cal N}_\text{2}(\nu, z,\alpha)\equiv &\pi \ell_3(\nu)  \sqrt{\nu ^2+1} z \left(\alpha ^2 \left(z^2-1\right)-1\right) (\nu  z+1) \left(-4 \alpha ^2 z^2+\nu  z \left(\alpha ^2
   \left(z^4+3\right)+z^2+3\right)+2\right)\\& \times \left(z^2 (\nu  z+1)+\alpha ^2 \left(-z^4+2 \nu  z^3-1\right)\right) \left(-6 \alpha ^2 z^2 \left(z^2-2 \nu 
   z-1\right)+z^2 (4 \nu  z+3)\right.\\&\left.+\alpha ^4 \left(z^2 \left(3 z^4-9 z^2+8 \nu  z-3\right)+1\right)-1\right)
\end{split}
\end{equation}

\noindent and

\begin{equation}
    \begin{split}
        \label{Cpalphadenominator}
        {\cal D}_\text{2}(\nu, z,\alpha)\equiv & \bigl(-\bigl(z^2 (2 \nu  z+3)\bigr)+\alpha ^2 \bigl(3 z^4-4 \nu  z^3+1\bigr)-1\bigr) \bigl(-2 z^2 (\nu  z+1)^2 \bigl(\nu  z^3-1\bigr) \bigl(\nu 
   \bigl(z^2-3\bigr) z+3 z^2-1\bigr)\\&-\alpha ^2 \bigl(z \bigl(2 \bigl(9 z^4-11 z^2+9\bigr) z+4 \nu ^4 \bigl(2 z^4+7 z^2-9\bigr) z^7+2 \nu ^3 \bigl(12
   z^6+65 z^4-38 z^2+21\bigr) z^4\\&+\nu  \bigl(86 z^6-72 z^4+70 z^2+4\bigr)+\nu ^2 \bigl(17 z^8+174 z^6-84 z^4+98 z^2+3\bigr) z\bigr)+2\bigr)+\alpha ^4
   z^2 \bigl(30 z^6\\&-26 z^4+74 z^2+2 \nu ^4 \bigl(z^2+3\bigr) \bigl(z^4-22 z^2+13\bigr) z^6+\nu ^2 (z-1) (z+1) \bigl(55 z^8+216 z^6-90 z^4\\&+184
   z^2+3\bigr)+4 \nu  \bigl(6 z^8+40 z^6-19 z^4+70 z^2-21\bigr) z+2 \nu ^3 \bigl(15 z^8-8 z^6-184 z^4+76 z^2-51\bigr) z^3\\&-14\bigr)+\alpha ^8 z^2
   \bigl(8 \nu ^4 \bigl(z^6-3 z^4-9 z^2+3\bigr) z^6+2 \nu  \bigl(14 z^2+3 \bigl(z^8+2 z^6-11 z^4+4 z^2-21\bigr) z^4-3\bigr) z\\&+3 \nu ^2 (z-1) (z+1)
   \bigl(z^{12}-8 z^{10}-35 z^8-29 z^4+8 z^2-1\bigr)+4 \bigl(3 z^{10}+3 z^8+12 z^6-4 z^4+z^2+1\bigr)\\&+4 \nu ^3 \bigl(-2 z^{10}+11 z^8+32 z^6-18 z^4+18
   z^2-9\bigr) z^3\bigr)+\alpha ^6 \bigl(z \bigl(20 z^3-6 \bigl(5 z^4+z^2+18\bigr) z^5+8 \nu ^4\\& \times \bigl(z^6-7 z^4-19 z^2+9\bigr) z^7+2 \nu ^3 \bigl(-2
   z^{10}+41 z^8+64 z^6-166 z^4+82 z^2-51\bigr) z^4+\nu ^2 \bigl(-27 z^{12}\\&+20 z^{10}+345 z^8-184 z^6+351 z^4-124 z^2+3\bigr) z-2 \nu  \bigl(12 z^{12}+51
   z^{10}-46 z^8+102 z^6-108 z^4\\&+23 z^2-2\bigr)-6 z\bigr)+2\bigr)\bigr).
    \end{split}
\end{equation}

$C_{p,z}$ is given by

\begin{equation}
     C_{p,z}(\nu,z,\alpha)=\frac{{\cal N}_\text{3}(\nu, z,\alpha)}{{\cal D}_\text{3}(\nu, z,\alpha)},
\end{equation}

\noindent where

\begin{equation}
    \begin{split}
        \label{Cpznumerator}
         {\cal N}_\text{3}(\nu, z,\alpha)\equiv &-2 \pi \ell_3(\nu)  \sqrt{\nu ^2+1} z \bigl(\alpha ^2 \bigl(z^2-1\bigr)-1\bigr) \bigl(-4 \alpha ^2 z^2+\nu  z \bigl(\alpha ^2
   \bigl(z^4+3\bigr)+z^2+3\bigr)+2\bigr) \bigl(z^2 (\nu  z+1)\\&+\alpha ^2 \bigl(-z^4+2 \nu  z^3-1\bigr)\bigr)
    \end{split}
\end{equation}

\noindent and

\begin{equation}
    \begin{split}
        \label{Cpzdenominator}
        {\cal D}_\text{3}(\nu, z,\alpha)\equiv &\bigl(-\bigl(z^2 (2 \nu  z+3)\bigr)+\alpha ^2 \bigl(3 z^4-4 \nu  z^3+1\bigr)-1\bigr) \bigl(z^2 \bigl(z \bigl(\nu  \bigl(z^2 (4 \nu  z+11)+6\bigr)+6
   z\bigr)+4\bigr)\\&+\alpha ^4 \bigl(6 z^8+16 \nu ^2 z^6+4 z^6+8 z^4-4 z^2-\nu  \bigl(z^8+20 z^6+2 z^4+12 z^2-3\bigr) z+2\bigr)+2 \alpha ^2 z\\& \times \bigl(8
   \nu ^2 z^5-2 \bigl(3 z^4+2 z^2+3\bigr) z+\nu  \bigl(-5 z^6+9 z^4+z^2+3\bigr)\bigr)+3 \nu  z+2\bigr)
.
    \end{split}
\end{equation}

$C_{V,\alpha}$ is given by

\begin{equation}
     C_{V,\alpha}(\nu,z,\alpha)=\frac{{\cal N}_\text{4}(\nu, z,\alpha)}{{\cal D}_\text{4}(\nu, z,\alpha)},
\end{equation}

\noindent where

\begin{equation}
    \begin{split}
        \label{CValphanumerator}
        {\cal N}_\text{4}(\nu, z,\alpha)\equiv& -2 \pi  \nu \ell_3(\nu)  \sqrt{\nu ^2+1} z \bigl(\alpha ^2 \bigl(z^2-1\bigr)-1\bigr) (\nu  z+1) \bigl(-4 \alpha ^2 z^2+\nu  z \bigl(\alpha ^2
   \bigl(z^4+3\bigr)+z^2+3\bigr)+2\bigr)\\& \times \bigl(z^2 (\nu  z+1)+\alpha ^2 \bigl(-z^4+2 \nu  z^3-1\bigr)\bigr) \bigl(z^2 (\nu  z+1)^3 \bigl(\nu 
   \bigl(z \bigl(\nu  \bigl(z^2 (4 \nu  z+3)+3\bigr)-3 z\bigr)-3\bigr)\\&+4 z\bigr)+2 \alpha ^8 z^2 (\nu -z) \bigl(-z^2+2 \nu  z+1\bigr) \bigl(z
   \bigl(8 \nu ^4 z^5+3 z^5+\nu  (z-1) (z+1) \bigl(z^4+3\bigr)-5 z^3\\&-\nu ^2 \bigl(z^2-3\bigr) \bigl(z^2+1\bigr)^2 z+\nu ^3 \bigl(3 z^6-9 z^4-7
   z^2-3\bigr) z^2+z\bigr)+1\bigr)+\alpha ^2 \bigl(28 \nu ^6 z^9+4 \bigl(3\\&-4 z^2\bigr) z^3+6 \nu ^5 \bigl(-3 z^4+13 z^2+2\bigr) z^6-2 \nu 
   \bigl(z^6+15 z^4-24 z^2+4\bigr) z^2+3 \nu ^4 \bigl(-15 z^4\\&+18 z^2+5\bigr) z^5+\nu ^3 \bigl(-28 z^8+17 z^6-9 z^4+3 z^2+1\bigr)-\nu ^2 \bigl(z^6+12
   z^4-51 z^2+22\bigr) z^3\bigr)\\&+\alpha ^4 z^2 \bigl(72 \nu ^6 z^7+26 z^5-36 z^3+3 \nu ^5 \bigl(z^6-31 z^4+43 z^2+3\bigr) z^4+\nu ^3 \bigl(38 z^8-48
   z^6\\&+22 z^4-36 z^2\bigr)+2 \nu ^2 \bigl(z^8+z^6-39 z^4+55 z^2-14\bigr) z+\nu  \bigl(4 z^8+31 z^6-111 z^4+73 z^2\\&-5\bigr)+6 \nu ^4 \bigl(6 z^6-25
   z^4+4 z^2-1\bigr) z^3+10 z\bigr)+\alpha ^6 \bigl(80 \nu ^6 z^9-20 \bigl(z^2-1\bigr)^2 z^5+12 \nu ^5 \bigl(z^6\\&-13 z^4+5 z^2-1\bigr) z^6-2 \nu 
   (z-1) (z+1) \bigl(z^8+7 z^6-36 z^4+19 z^2+1\bigr) z^2-3 \nu ^4 \bigl(3 z^8\\&-36 z^6+38 z^4+4 z^2+7\bigr) z^5-\nu ^2 (z-1) (z+1) \bigl(z^8-5 z^6-39
   z^4+73 z^2-14\bigr) z^3\\&+\nu ^3 \bigl(-24 z^{12}+49 z^{10}-11 z^8+60 z^6-12 z^4+3 z^2-1\bigr)\bigr)\bigr)
    \end{split}
\end{equation}

\noindent and

\begin{align}
        \label{CValphadenominator}
      {\cal D}_\text{4}(\nu, z,\alpha)\equiv &\bigl((2 z \nu +3) z^2+\alpha ^2 \bigl(-3 z^4+4 \nu  z^3-1\bigr)+1\bigr) \bigl(12 \alpha ^{12} \nu ^6 z^{24}-\alpha ^{10} \nu ^5 \bigl(\bigl(70 \nu
   ^2-8\bigr) \alpha ^2+35 \nu ^2\\& \nonumber+2\bigr) z^{23}+\alpha ^8 \nu ^4 \bigl(39 \bigl(2 \alpha ^2+1\bigr)^2 \nu ^4+\bigl(-146 \alpha ^4-147 \alpha
   ^2+2\bigr) \nu ^2-4 \alpha ^2 \bigl(17 \alpha ^2+4\bigr)\bigr) z^{22}\\& \nonumber+\alpha ^6 \nu ^3 \bigl(-20 \bigl(2 \alpha ^2+1\bigr)^3 \nu ^6+25 \alpha ^2
   \bigl(22 \alpha ^4+39 \alpha ^2+14\bigr) \nu ^4+2 \alpha ^2 \bigl(155 \alpha ^4+64 \alpha ^2+18\bigr) \nu ^2\\& \nonumber-8 \alpha ^4 \bigl(10 \alpha
   ^2+11\bigr)\bigr) z^{21}+\alpha ^4 \nu ^2 \bigl(4 \bigl(2 \alpha ^2+1\bigr)^4 \nu ^8-3 \bigl(71 \alpha ^2+104\bigr) \bigl(2 \alpha ^3+\alpha
   \bigr)^2 \nu ^6+\alpha ^2\\& \nonumber \times \bigl(-722 \alpha ^6+123 \alpha ^4+439 \alpha ^2-14\bigr) \nu ^4+2 \alpha ^4 \bigl(189 \alpha ^4+425 \alpha ^2+124\bigr)
   \nu ^2+24 \alpha ^6 \bigl(4 \alpha ^2\\& \nonumber-5\bigr)\bigr) z^{20}+2 \alpha ^4 \nu  \bigl(20 \bigl(2 \alpha ^2+1\bigr)^3 \bigl(2 \alpha ^2+3\bigr) \nu
   ^8+\alpha ^2 \bigl(824 \alpha ^6+90 \alpha ^4-1047 \alpha ^2-443\bigr) \nu ^6\\& \nonumber-\alpha ^2 \times \bigl(361 \alpha ^6+1102 \alpha ^4+480 \alpha ^2+94\bigr)
   \nu ^4+2 \alpha ^4 \bigl(-161 \alpha ^4+120 \alpha ^2+140\bigr) \nu ^2+12 \alpha ^6\\& \nonumber \times \bigl(9 \alpha ^2-2\bigr)\bigr) z^{19}-\alpha ^2 \bigl(-96
   \alpha ^{10}+4 \bigl(365 \alpha ^4+212 \alpha ^2-136\bigr) \nu ^2 \alpha ^6+2 \bigl(-721 \alpha ^6-7 \alpha ^4\\& \nonumber+1042 \alpha ^2+352\bigr) \nu ^4
   \alpha ^4-\bigl(1152 \alpha ^8+4612 \alpha ^6+1799 \alpha ^4-416 \alpha ^2+38\bigr) \nu ^6 \alpha ^2+4 \bigl(2 \alpha ^2+1\bigr)^4\\& \nonumber \times \bigl(3 \alpha
   ^2+4\bigr) \nu ^{10}+\bigl(2 \alpha ^3+\alpha \bigr)^2 \bigl(613 \alpha ^4+70 \alpha ^2-506\bigr) \nu ^8\bigr) z^{18}+2 \alpha ^2 \nu  \bigl(2
   \bigl(2 \alpha ^2+1\bigr)^3 \bigl(58 \alpha ^4\\& \nonumber+38 \alpha ^2-21\bigr) \nu ^8-\alpha ^2 \bigl( \bigl(440 \alpha ^6+2113 \alpha
   ^4+919 \alpha ^2-362\bigr) \nu ^6+2 \alpha ^2 \bigl(-448 \alpha ^8-427 \alpha ^6+568 \alpha ^4\\& \nonumber+310 \alpha ^2+69\bigr) \nu ^4+2 \alpha ^4 \bigl(904
   \alpha ^6+1373 \alpha ^4-9 \alpha ^2-288\bigr) \nu ^2-48 \alpha ^6 \bigl(11 \alpha ^4+13 \alpha ^2-2\bigr)\bigr) z^{17}\\& \nonumber-\bigl(96 \bigl(3 \alpha
   ^2+5\bigr) \alpha ^{10}-4 \bigl(1130 \alpha ^6+2117 \alpha ^4+788 \alpha ^2-228\bigr) \nu ^2 \alpha ^6+2 \bigl(2182 \alpha ^8+6031 \alpha ^6\\& \nonumber+3609
   \alpha ^4 -328 \alpha ^2-364\bigr) \nu ^4 \alpha ^4+\bigl(-240 \alpha ^{10}-26 \alpha ^8+5941 \alpha ^6+3310 \alpha ^4-194 \alpha ^2+26\bigr) \nu ^6
   \alpha ^2\\& \nonumber+4 \bigl(\alpha ^2+1\bigr) \bigl(2 \alpha ^2+1\bigr)^4 \bigl(9 \alpha ^2+1\bigr) \nu ^{10}-\bigl(2 \alpha ^3+\alpha \bigr)^2 \bigl(391
   \alpha ^6+2150 \alpha ^4+1444 \alpha ^2-136\bigr) \nu ^8\bigr) z^{16}\\& \nonumber+\nu  \bigl(-8 \bigl(2 \alpha ^2+1\bigr)^3 \bigl(12 \alpha ^6+59 \alpha ^4+49
   \alpha ^2+4\bigr) \nu ^8+\alpha ^2 \bigl(2 \alpha ^2+1\bigr) \bigl(1034 \alpha ^8+1998 \alpha ^6+5766 \alpha ^4\\& \nonumber+3604 \alpha ^2-31\bigr) \nu ^6+2
   \alpha ^2 \bigl(2 \alpha ^2 \bigl(412 \alpha ^8+2544 \alpha ^6+3319 \alpha ^4+949 \alpha ^2+77\bigr)-61\bigr) \nu ^4-4 \alpha ^4 \bigl(2072 \alpha
   ^8\\& \nonumber+5404 \alpha ^6+4633 \alpha ^4+804 \alpha ^2-244\bigr) \nu ^2+16 \alpha ^6 \bigl(160 \alpha ^6+336 \alpha ^4+197 \alpha ^2-18\bigr)\bigr)
   z^{15}+\bigl(12\\& \nonumber \times\bigl(\alpha ^2+1\bigr)^2 \bigl(2 \alpha ^2+1\bigr)^4 \nu ^{10}-\bigl(2 \alpha ^2+1\bigr)^2 \bigl(571 \alpha ^8+948 \alpha ^6+1780
   \alpha ^4+1314 \alpha ^2+97\bigr) \nu ^8+\alpha ^2\\& \nonumber \times \bigl(2428 \alpha ^{10}+4052 \alpha ^8-1635 \alpha ^6+1290 \alpha ^4+1828 \alpha ^2-121\bigr) \nu
   ^6+2 \alpha ^2 \bigl(3806 \alpha ^{10}+12187 \alpha ^8\\& \nonumber+16656 \alpha ^6+9115 \alpha ^4+1330 \alpha ^2-128\bigr) \nu ^4-4 \alpha ^4 \bigl(1852 \alpha
   ^8+5346 \alpha ^6+5119 \alpha ^4+1528 \alpha ^2\\& \nonumber-168\bigr) \nu ^2+48 \alpha ^8 \bigl(10 \alpha ^4+26 \alpha ^2+21\bigr)\bigr) z^{14}+\nu  \bigl(4
   \bigl(\alpha ^2+1\bigr) \bigl(2 \alpha ^2+1\bigr)^3 \bigl(27 \alpha ^4+15 \alpha ^2+16\bigr) \nu ^8-\bigl(2 \alpha ^2\\& \nonumber+1\bigr) \bigl(1634 \alpha
   ^{10}+5374 \alpha ^8+3702 \alpha ^6+2776 \alpha ^4+1893 \alpha ^2+122\bigr) \nu ^6-2 \alpha ^2 \bigl(1306 \alpha ^{10}+3406 \alpha ^8+7550 \alpha
   ^6\\& \nonumber +8714 \alpha ^4+3203 \alpha ^2+532\bigr) \nu ^4+4 \alpha ^2 \bigl(2540 \alpha ^{10}+9156 \alpha ^8+12001 \alpha ^6+7208 \alpha ^4+1452 \alpha
   ^2-74\bigr) \nu ^2\\& \nonumber-32 \alpha ^4 \bigl(109 \alpha ^8+333 \alpha ^6+348 \alpha ^4+137 \alpha ^2-6\bigr)\bigr) z^{13}+\bigl(\bigl(2 \alpha
   ^2+1\bigr)^2 \bigl(417 \alpha ^8+1468 \alpha ^6+1056 \alpha ^4\\& \nonumber+246 \alpha ^2+137\bigr) \nu ^8-\bigl(1004 \alpha ^{12}+8184 \alpha ^{10}+15507 \alpha
   ^8+6554 \alpha ^6+1054 \alpha ^4+663 \alpha ^2-23\bigr) \nu ^6-2 \alpha ^2\\& \nonumber \times \bigl(3380 \alpha ^{10}+14197 \alpha ^8+21537 \alpha ^6+18393 \alpha
   ^4+8076 \alpha ^2+1341\bigr) \nu ^4+4 \alpha ^2 \bigl(1938 \alpha ^{10}+7307 \alpha ^8\\& \nonumber+10151 \alpha ^6+6370 \alpha ^4+1568 \alpha ^2-46\bigr) \nu
   ^2-48 \alpha ^6 \bigl(14 \alpha ^6+40 \alpha ^4+47 \alpha ^2+24\bigr)\bigr) z^{12}-2 \nu  \bigl(24 \bigl(2 \alpha ^2+1\bigr)^3\\& \nonumber \times \bigl(\alpha
   ^3+\alpha \bigr)^2 \nu ^8-\bigl(2 \alpha ^2+1\bigr) \bigl(328 \alpha ^{10}+1589 \alpha ^8+3235 \alpha ^6+1912 \alpha ^4+301 \alpha ^2+95\bigr) \nu
   ^6-\bigl(574 \alpha ^{12}\\& \nonumber+3182 \alpha ^{10}+3340 \alpha ^8+2260 \alpha ^6+2791 \alpha ^4+1012 \alpha ^2+170\bigr) \nu ^4+2 \alpha ^2 \bigl(1998
   \alpha ^{10}+8724 \alpha ^8+14616 \alpha ^6\\& \nonumber+11995 \alpha ^4+5210 \alpha ^2+936\bigr) \nu ^2-4 \alpha ^2 \bigl(362 \alpha ^{10}+1504 \alpha ^8+2221
   \alpha ^6+1490 \alpha ^4+435 \alpha ^2-6\bigr)\bigr) z^{11}+\\& \nonumber\bigl(-\bigl(\bigl(\alpha ^2+1\bigr) \bigl(2 \alpha ^2+1\bigr)^2 \bigl(135 \alpha
   ^6+287 \alpha ^4+295 \alpha ^2+15\bigr) \nu ^8\bigr)+\bigl(912 \alpha ^{12}+2866 \alpha ^{10}+6643 \alpha ^8+11770 \alpha ^6\\& \nonumber+6972 \alpha ^4+1408
   \alpha ^2+233\bigr) \nu ^6+2 \bigl(1788 \alpha ^{12}+8089 \alpha ^{10}+15578 \alpha ^8+13774 \alpha ^6+7417 \alpha ^4+2382 \alpha ^2\\& \nonumber+334\bigr) \nu
   ^4-4 \alpha ^2 \bigl(1214 \alpha ^{10}+6004 \alpha ^8+10726 \alpha ^6+9259 \alpha ^4+4049 \alpha ^2+788\bigr) \nu ^2+32 \alpha ^4 \bigl(\alpha
   ^2+1\bigr) \bigl(17 \alpha ^6\\& \nonumber+53 \alpha ^4+44 \alpha ^2+24\bigr)\bigr) z^{10}-2 \nu  \bigl(\bigl(2 \alpha ^2+1\bigr) \bigl(156 \alpha ^{10}+475
   \alpha ^8+405 \alpha ^6+470 \alpha ^4+315 \alpha ^2+19\bigr) \nu ^6+2\\& \nonumber \times \bigl(60 \alpha ^{12}-31 \alpha ^{10}+332 \alpha ^8+100 \alpha ^6-637 \alpha
   ^4-226 \alpha ^2-45\bigr) \nu ^4-2 \bigl(992 \alpha ^{12}+5018 \alpha ^{10}+9806 \alpha ^8\\& \nonumber+9711 \alpha ^6+5136 \alpha ^4+1501 \alpha ^2+197\bigr)
   \nu ^2+8 \alpha ^2 \bigl(94 \alpha ^{10}+530 \alpha ^8+1022 \alpha ^6+926 \alpha ^4+423 \alpha ^2+92\bigr)\bigr) z^9\\& \nonumber+\bigl(9 \bigl(2 \alpha ^4+3
   \alpha ^2+1\bigr)^2 \bigl(5 \alpha ^4-4 \alpha ^2-1\bigr) \nu ^8-\bigl(300 \alpha ^{12}+2026 \alpha ^{10}+2983 \alpha ^8+690 \alpha ^6+362 \alpha
   ^4+454 \alpha ^2\\& \nonumber+19\bigr) \nu ^6-2 \bigl(730 \alpha ^{12}+3149 \alpha ^{10}+5333 \alpha ^8+5634 \alpha ^6+2662 \alpha ^4+564 \alpha ^2+36\bigr) \nu
   ^4+4 \bigl(446 \alpha ^{12}+2859 \alpha ^{10}\\& \nonumber  +6326 \alpha ^8+6766 \alpha ^6+3836 \alpha ^4+1117 \alpha ^2+148\bigr) \nu ^2-96 \bigl(\alpha ^3+\alpha
   \bigr)^2 \bigl(\alpha ^6+11 \alpha ^4+6 \alpha ^2+3\bigr)\bigr) z^8+\nu\\& \nonumber \times  \bigl(\bigl(\alpha ^2+1\bigr) \bigl(2 \alpha ^2+1\bigr) \bigl(81 \alpha
   ^8+249 \alpha ^6+49 \alpha ^4-181 \alpha ^2-30\bigr) \nu ^6+2 \bigl(3 \bigl(36 \alpha ^6+107 \alpha ^4-46 \alpha ^2-58\bigr) \alpha ^6\\& \nonumber+338 \alpha
   ^4+23 \alpha ^2+14\bigr) \nu ^4-4 \bigl(248 \alpha ^{12}+1492 \alpha ^{10}+3257 \alpha ^8+3602 \alpha ^6+2195 \alpha ^4+657 \alpha ^2+78\bigr) \nu
   ^2\\& \nonumber+8 \bigl(32 \alpha ^{12}+416 \alpha ^{10}+1092 \alpha ^8+1262 \alpha ^6+745 \alpha ^4+222 \alpha ^2+32\bigr)\bigr) z^7+\bigl(\bigl(6 \alpha
   ^{12}+15 \alpha ^{10}+45 \alpha ^8\\& \nonumber+118 \alpha ^6-316 \alpha ^4-311 \alpha ^2-37\bigr) \nu ^6+2 \bigl(116 \alpha ^{12}+521 \alpha ^{10}+932 \alpha
   ^8+543 \alpha ^6+388 \alpha ^4+126 \alpha ^2+20\bigr) \nu ^4\\& \nonumber-4 \bigl(92 \alpha ^{12}+794 \alpha ^{10}+2015 \alpha ^8+2420 \alpha ^6+1558 \alpha
   ^4+529 \alpha ^2+71\bigr) \nu ^2-48 \bigl(\alpha ^2+1\bigr)^3 \bigl(2 \alpha ^6-8 \alpha ^4-3 \alpha ^2\\& \nonumber-1\bigr)\bigr) z^6+\nu  \bigl(-3
   \bigl(\alpha ^3+\alpha \bigr)^2 \bigl(6 \alpha ^6-25 \alpha ^4-16 \alpha ^2-1\bigr) \nu ^6-2 \bigl(9 \alpha ^{12}+26 \alpha ^{10}+200 \alpha ^8+234
   \alpha ^6+69 \alpha ^4\\& \nonumber+88 \alpha ^2+10\bigr) \nu ^4+4 \bigl(64 \alpha ^{12}+314 \alpha ^{10}+585 \alpha ^8+514 \alpha ^6+225 \alpha ^4+57 \alpha
   ^2+5\bigr) \nu ^2+16 \bigl(\alpha ^2+1\bigr) \bigl(10 \alpha ^{10}\\& \nonumber-24 \alpha ^8-94 \alpha ^6-96 \alpha ^4-44 \alpha ^2-7\bigr)\bigr)
   z^5-\bigl(\alpha ^2 \bigl(\alpha ^2+1\bigr) \bigl(18 \alpha ^8+33 \alpha ^6-153 \alpha ^4-143 \alpha ^2-7\bigr) \nu ^6\\& \nonumber+2 \bigl(41 \alpha ^{12}+152
   \alpha ^{10}+235 \alpha ^8+253 \alpha ^6+76 \alpha ^4+27 \alpha ^2+2\bigr) \nu ^4+4 \bigl(\alpha ^2+1\bigr) \bigl(18 \alpha ^{10}-53 \alpha ^8-154
   \alpha ^6\\& \nonumber-122 \alpha ^4-30 \alpha ^2-1\bigr) \nu ^2-16 \bigl(\alpha ^2+1\bigr)^4 \bigl(2 \alpha ^4-4 \alpha ^2-1\bigr)\bigr) z^4+2 \alpha ^2 \nu 
   \bigl(\bigl(3 \alpha ^{10}+16 \alpha ^6+62 \alpha ^4+61 \alpha ^2\\& \nonumber+2\bigr) \nu ^4-2 \bigl(\alpha ^2+1\bigr) \bigl(\alpha ^8+19 \alpha ^6+31 \alpha
   ^4+24 \alpha ^2+3\bigr) \nu ^2-4 \bigl(\alpha ^2+1\bigr)^2 \bigl(5 \alpha ^2 \bigl(\alpha ^4-2\bigr)-4\bigr)\bigr) z^3+2 \alpha ^2 \nu ^2\\& \nonumber \times
   \bigl(-3 \bigl(\alpha ^2-1\bigr) \bigl(\alpha ^3+\alpha \bigr)^2 \nu ^4+\alpha ^2 \bigl(3 \alpha ^8+4 \alpha ^6+4 \alpha ^4+8 \alpha ^2+17\bigr)
   \nu ^2+2 \bigl(\alpha ^2+1\bigr)^2 \bigl(3 \alpha ^6-2 \alpha ^4-3 \alpha ^2\\& \nonumber-1\bigr)\bigr) z^2+4 \alpha ^4 \bigl(\alpha ^4-1\bigr) \nu ^3
   \bigl(\alpha ^2-2 \nu ^2+1\bigr) z-4 \alpha ^4 \bigl(\alpha ^4-1\bigr) \nu ^4\bigr).
\end{align}

$C_{V,z}$ is given by

\begin{equation}
     C_{V,z}(\nu,z,\alpha)=\frac{{\cal N}_\text{5}(\nu, z,\alpha)}{{\cal D}_\text{5}(\nu, z,\alpha)},
\end{equation}

\noindent where

\begin{equation}
    \begin{split}
        \label{CVznumerator}
        {\cal N}_\text{5}(\nu, z,\alpha)\equiv &-\pi \ell_3(\nu)  \nu  \sqrt{\nu ^2+1} z \bigl(\alpha ^2 \bigl(z^2-1\bigr)-1\bigr) (\nu  z+1) \bigl(-4 \alpha ^2 z^2+\nu  z \bigl(\alpha ^2
   \bigl(z^4+3\bigr)+z^2\\&+3\bigr)+2\bigr) \bigl(z^2 (\nu  z+1)+\alpha ^2 \bigl(-z^4+2 \nu  z^3-1\bigr)\bigr) \bigl(4 \bigl(2 \alpha ^2+1\bigr)^2 \nu
   ^5 z^6+2 \bigl(z^2-1\bigr) z\\& \times \bigl(\nu -\alpha ^2 \nu  \bigl(z^2-1\bigr)\bigr)^2-2 \bigl(z^2-1\bigr) z \bigl(\alpha ^2
   \bigl(z^2-1\bigr)-1\bigr)^2+\nu ^3 \bigl(-\alpha ^2 \bigl(3 z^4+1\bigr)+3 z^2\\&+1\bigr)^2-\nu  \bigl(z^4-6 z^2+1\bigr) \bigl(\alpha ^2
   \bigl(z^2-1\bigr)-1\bigr)^2-4 \bigl(2 \alpha ^2+1\bigr) \nu ^4 z^3 \bigl(\alpha ^2 \bigl(3 z^4+1\bigr)\\&-3 z^2-1\bigr)\bigr)
    \end{split}
\end{equation}

\noindent and

\begin{align}
        \label{CVzdenominator}
        {\cal D}_\text{5}(\nu, z,\alpha)\equiv &\bigl((2 z \nu +3) z^2+\alpha ^2 \bigl(-3 z^4+4 \nu  z^3-1\bigr)+1\bigr) \bigl(12 \alpha ^{12} \nu ^6 z^{24}-\alpha ^{10} \nu ^5 \bigl(\bigl(70 \nu
   ^2-8\bigr) \alpha ^2\\& \nonumber+35 \nu ^2+2\bigr) z^{23}+\alpha ^8 \nu ^4 \bigl(39 \bigl(2 \alpha ^2+1\bigr)^2 \nu ^4+\bigl(-146 \alpha ^4-147 \alpha
   ^2+2\bigr) \nu ^2-4 \alpha ^2 \bigl(17 \alpha ^2\\& \nonumber+4\bigr)\bigr) z^{22}+\alpha ^6 \nu ^3 \bigl(-20 \bigl(2 \alpha ^2+1\bigr)^3 \nu ^6+25 \alpha ^2
   \bigl(22 \alpha ^4+39 \alpha ^2+14\bigr) \nu ^4+2 \alpha ^2 \bigl(155 \alpha ^4\\& \nonumber+64 \alpha ^2+18\bigr) \nu ^2-8 \alpha ^4 \bigl(10 \alpha
   ^2+11\bigr)\bigr) z^{21}+\alpha ^4 \nu ^2 \bigl(4 \bigl(2 \alpha ^2+1\bigr)^4 \nu ^8-3 \bigl(71 \alpha ^2+104\bigr) \bigl(2 \alpha ^3\\& \nonumber+\alpha
   \bigr)^2 \nu ^6+\alpha ^2 \bigl(-722 \alpha ^6+123 \alpha ^4+439 \alpha ^2-14\bigr) \nu ^4+2 \alpha ^4 \bigl(189 \alpha ^4+425 \alpha ^2+124\bigr)
   \nu ^2\\& \nonumber+24 \alpha ^6 \bigl(4 \alpha ^2-5\bigr)\bigr) z^{20}+2 \alpha ^4 \nu  \bigl(20 \bigl(2 \alpha ^2+1\bigr)^3 \bigl(2 \alpha ^2+3\bigr) \nu
   ^8+\alpha ^2 \bigl(824 \alpha ^6+90 \alpha ^4-1047 \alpha ^2\\& \nonumber-443\bigr) \nu ^6-\alpha ^2 \bigl(361 \alpha ^6+1102 \alpha ^4+480 \alpha ^2+94\bigr) \nu
   ^4+2 \alpha ^4 \bigl(-161 \alpha ^4+120 \alpha ^2+140\bigr) \nu ^2\\& \nonumber+12 \alpha ^6 \bigl(9 \alpha ^2-2\bigr)\bigr) z^{19}-\alpha ^2 \bigl(-96 \alpha
   ^{10}+4 \bigl(365 \alpha ^4+212 \alpha ^2-136\bigr) \nu ^2 \alpha ^6+2 \bigl(-721 \alpha ^6\\& \nonumber-7 \alpha ^4+1042 \alpha ^2+352\bigr) \nu ^4 \alpha
   ^4-\bigl(1152 \alpha ^8+4612 \alpha ^6+1799 \alpha ^4-416 \alpha ^2+38\bigr) \nu ^6 \alpha ^2\\& \nonumber+4 \bigl(2 \alpha ^2+1\bigr)^4 \bigl(3 \alpha ^2+4\bigr)
   \nu ^{10}+\bigl(2 \alpha ^3+\alpha \bigr)^2 \bigl(613 \alpha ^4+70 \alpha ^2-506\bigr) \nu ^8\bigr) z^{18}+2 \alpha ^2 \nu  \bigl(2 \bigl(2 \alpha
   ^2\\& \nonumber+1\bigr)^3 \bigl(58 \alpha ^4+38 \alpha ^2-21\bigr) \nu ^8-\alpha ^2 \bigl(2 \alpha ^2+1\bigr) \bigl(440 \alpha ^6+2113 \alpha ^4+919 \alpha
   ^2-362\bigr) \nu ^6+2 \alpha ^2 \\& \nonumber \times \bigl(-448 \alpha ^8-427 \alpha ^6+568 \alpha ^4+310 \alpha ^2+69\bigr) \nu ^4+2 \alpha ^4 \bigl(904 \alpha ^6+1373
   \alpha ^4-9 \alpha ^2-288\bigr) \nu ^2\\& \nonumber-48 \alpha ^6 \bigl(11 \alpha ^4+13 \alpha ^2-2\bigr)\bigr) z^{17}-\bigl(96 \bigl(3 \alpha ^2+5\bigr) \alpha
   ^{10}-4 \bigl(1130 \alpha ^6+2117 \alpha ^4+788 \alpha ^2-228\bigr)\\& \nonumber \times \nu ^2 \alpha ^6+2 \bigl(2182 \alpha ^8+6031 \alpha ^6+3609 \alpha ^4-328 \alpha
   ^2-364\bigr) \nu ^4 \alpha ^4+\bigl(-240 \alpha ^{10}-26 \alpha ^8\\& \nonumber+5941 \alpha ^6+3310 \alpha ^4-194 \alpha ^2+26\bigr) \nu ^6 \alpha ^2+4
   \bigl(\alpha ^2+1\bigr) \bigl(2 \alpha ^2+1\bigr)^4 \bigl(9 \alpha ^2+1\bigr) \nu ^{10}\\& \nonumber-\bigl(2 \alpha ^3+\alpha \bigr)^2 \bigl(391 \alpha ^6+2150
   \alpha ^4+1444 \alpha ^2-136\bigr) \nu ^8\bigr) z^{16}+\nu  \bigl(-8 \bigl(2 \alpha ^2+1\bigr)^3 \bigl(12 \alpha ^6+59 \alpha ^4\\& \nonumber+49 \alpha
   ^2+4\bigr) \nu ^8+\alpha ^2 \bigl(2 \alpha ^2+1\bigr) \bigl(1034 \alpha ^8+1998 \alpha ^6+5766 \alpha ^4+3604 \alpha ^2-31\bigr) \nu ^6+2 \alpha ^2\\& \nonumber \times
   \bigl(2 \alpha ^2 \bigl(412 \alpha ^8+2544 \alpha ^6+3319 \alpha ^4+949 \alpha ^2+77\bigr)-61\bigr) \nu ^4-4 \alpha ^4 \bigl(2072 \alpha ^8+5404
   \alpha ^6\\& \nonumber+4633 \alpha ^4+804 \alpha ^2-244\bigr) \nu ^2+16 \alpha ^6 \bigl(160 \alpha ^6+336 \alpha ^4+197 \alpha ^2-18\bigr)\bigr) z^{15}+\bigl(12
   \bigl(\alpha ^2\\& \nonumber+1\bigr)^2 \bigl(2 \alpha ^2+1\bigr)^4 \nu ^{10}-\bigl(2 \alpha ^2+1\bigr)^2 \bigl(571 \alpha ^8+948 \alpha ^6+1780 \alpha ^4+1314
   \alpha ^2+97\bigr) \nu ^8\\& \nonumber+\alpha ^2 \bigl(2428 \alpha ^{10}+4052 \alpha ^8-1635 \alpha ^6+1290 \alpha ^4+1828 \alpha ^2-121\bigr) \nu ^6+2 \alpha ^2
   \bigl(3806 \alpha ^{10}\\& \nonumber+12187 \alpha ^8+16656 \alpha ^6+9115 \alpha ^4+1330 \alpha ^2-128\bigr) \nu ^4-4 \alpha ^4 \bigl(1852 \alpha ^8+5346 \alpha
   ^6+5119 \alpha ^4\\& \nonumber+1528 \alpha ^2-168\bigr) \nu ^2+48 \alpha ^8 \bigl(10 \alpha ^4+26 \alpha ^2+21\bigr)\bigr) z^{14}+\nu  \bigl(4 \bigl(\alpha
   ^2+1\bigr) \bigl(2 \alpha ^2+1\bigr)^3 \bigl(27 \alpha ^4\\& \nonumber+15 \alpha ^2+16\bigr) \nu ^8-\bigl(2 \alpha ^2+1\bigr) \bigl(1634 \alpha ^{10}+5374 \alpha
   ^8+3702 \alpha ^6+2776 \alpha ^4+1893 \alpha ^2+122\bigr) \nu ^6\\& \nonumber-2 \alpha ^2 \bigl(1306 \alpha ^{10}+3406 \alpha ^8+7550 \alpha ^6+8714 \alpha ^4+3203
   \alpha ^2+532\bigr) \nu ^4+4 \alpha ^2 \bigl(2540 \alpha ^{10}\\& \nonumber+9156 \alpha ^8+12001 \alpha ^6+7208 \alpha ^4+1452 \alpha ^2-74\bigr) \nu ^2-32 \alpha
   ^4 \bigl(109 \alpha ^8+333 \alpha ^6+348 \alpha ^4\\& \nonumber+137 \alpha ^2-6\bigr)\bigr) z^{13}+\bigl(\bigl(2 \alpha ^2+1\bigr)^2 \bigl(417 \alpha ^8+1468
   \alpha ^6+1056 \alpha ^4+246 \alpha ^2+137\bigr) \nu ^8\\& \nonumber-\bigl(1004 \alpha ^{12}+8184 \alpha ^{10}+15507 \alpha ^8+6554 \alpha ^6+1054 \alpha ^4+663
   \alpha ^2-23\bigr) \nu ^6-2 \alpha ^2\\& \nonumber \times \bigl(3380 \alpha ^{10}+14197 \alpha ^8+21537 \alpha ^6+18393 \alpha ^4+8076 \alpha ^2+1341\bigr) \nu ^4+4
   \alpha ^2 \bigl(1938 \alpha ^{10}\\& \nonumber+7307 \alpha ^8+10151 \alpha ^6+6370 \alpha ^4+1568 \alpha ^2-46\bigr) \nu ^2-48 \alpha ^6 \bigl(14 \alpha ^6+40
   \alpha ^4+47 \alpha ^2+24\bigr)\bigr) z^{12}\\& \nonumber-2 \nu  \bigl(24 \bigl(2 \alpha ^2+1\bigr)^3 \bigl(\alpha ^3+\alpha \bigr)^2 \nu ^8-\bigl(2 \alpha
   ^2+1\bigr) \bigl(328 \alpha ^{10}+1589 \alpha ^8+3235 \alpha ^6+1912 \alpha ^4+301 \alpha ^2\\& \nonumber+95\bigr) \nu ^6-\bigl(574 \alpha ^{12}+3182 \alpha
   ^{10}+3340 \alpha ^8+2260 \alpha ^6+2791 \alpha ^4+1012 \alpha ^2+170\bigr) \nu ^4+2 \alpha ^2\\& \nonumber \times \bigl(1998 \alpha ^{10}+8724 \alpha ^8+14616 \alpha
   ^6+11995 \alpha ^4+5210 \alpha ^2+936\bigr) \nu ^2-4 \alpha ^2 \bigl(362 \alpha ^{10}+1504 \alpha ^8\\& \nonumber+2221 \alpha ^6+1490 \alpha ^4+435 \alpha
   ^2-6\bigr)\bigr) z^{11}+\bigl(-\bigl(\bigl(\alpha ^2+1\bigr) \bigl(2 \alpha ^2+1\bigr)^2 \bigl(135 \alpha ^6+287 \alpha ^4+295 \alpha ^2\\& \nonumber+15\bigr)
   \nu ^8\bigr)+\bigl(912 \alpha ^{12}+2866 \alpha ^{10}+6643 \alpha ^8+11770 \alpha ^6+6972 \alpha ^4+1408 \alpha ^2+233\bigr) \nu ^6\\& \nonumber+2 \bigl(1788
   \alpha ^{12}+8089 \alpha ^{10}+15578 \alpha ^8+13774 \alpha ^6+7417 \alpha ^4+2382 \alpha ^2+334\bigr) \nu ^4-4 \alpha ^2 \bigl(1214 \alpha ^{10}\\& \nonumber+6004
   \alpha ^8+10726 \alpha ^6+9259 \alpha ^4+4049 \alpha ^2+788\bigr) \nu ^2+32 \alpha ^4 \bigl(\alpha ^2+1\bigr) \bigl(17 \alpha ^6+53 \alpha ^4+44
   \alpha ^2\\& \nonumber+24\bigr)\bigr) z^{10}-2 \nu  \bigl(\bigl(2 \alpha ^2+1\bigr) \bigl(156 \alpha ^{10}+475 \alpha ^8+405 \alpha ^6+470 \alpha ^4+315 \alpha
   ^2+19\bigr) \nu ^6+2 \bigl(60 \alpha ^{12}\\& \nonumber-31 \alpha ^{10}+332 \alpha ^8+100 \alpha ^6-637 \alpha ^4-226 \alpha ^2-45\bigr) \nu ^4-2 \bigl(992 \alpha
   ^{12}+5018 \alpha ^{10}+9806 \alpha ^8\\& \nonumber+9711 \alpha ^6+5136 \alpha ^4+1501 \alpha ^2+197\bigr) \nu ^2+8 \alpha ^2 \bigl(94 \alpha ^{10}+530 \alpha
   ^8+1022 \alpha ^6+926 \alpha ^4+423 \alpha ^2\\& \nonumber+92\bigr)\bigr) z^9+\bigl(9 \bigl(2 \alpha ^4+3 \alpha ^2+1\bigr)^2 \bigl(5 \alpha ^4-4 \alpha
   ^2-1\bigr) \nu ^8-\bigl(300 \alpha ^{12}+2026 \alpha ^{10}+2983 \alpha ^8+690 \alpha ^6\\& \nonumber+362 \alpha ^4+454 \alpha ^2+19\bigr) \nu ^6-2 \bigl(730 \alpha
   ^{12}+3149 \alpha ^{10}+5333 \alpha ^8+5634 \alpha ^6+2662 \alpha ^4+564 \alpha ^2+36\bigr) \nu ^4\\& \nonumber+4 \bigl(446 \alpha ^{12}+2859 \alpha ^{10}+6326
   \alpha ^8+6766 \alpha ^6+3836 \alpha ^4+1117 \alpha ^2+148\bigr) \nu ^2-96 \bigl(\alpha ^3+\alpha \bigr)^2 \\& \nonumber \times \bigl(\alpha ^6+11 \alpha ^4+6 \alpha
   ^2+3\bigr)\bigr) z^8+\nu  \bigl(\bigl(\alpha ^2+1\bigr) \bigl(2 \alpha ^2+1\bigr) \bigl(81 \alpha ^8+249 \alpha ^6+49 \alpha ^4-181 \alpha
   ^2-30\bigr) \nu ^6\\& \nonumber+2 \bigl(3 \bigl(36 \alpha ^6+107 \alpha ^4-46 \alpha ^2-58\bigr) \alpha ^6+338 \alpha ^4+23 \alpha ^2+14\bigr) \nu ^4-4 \bigl(248
   \alpha ^{12}+1492 \alpha ^{10}\\& \nonumber+3257 \alpha ^8+3602 \alpha ^6+2195 \alpha ^4+657 \alpha ^2+78\bigr) \nu ^2+8 \bigl(32 \alpha ^{12}+416 \alpha ^{10}+1092
   \alpha ^8+1262 \alpha ^6\\& \nonumber+745 \alpha ^4+222 \alpha ^2+32\bigr)\bigr) z^7+\bigl(\bigl(6 \alpha ^{12}+15 \alpha ^{10}+45 \alpha ^8+118 \alpha ^6-316
   \alpha ^4-311 \alpha ^2-37\bigr) \nu ^6\\& \nonumber+2 \bigl(116 \alpha ^{12}+521 \alpha ^{10}+932 \alpha ^8+543 \alpha ^6+388 \alpha ^4+126 \alpha ^2+20\bigr) \nu
   ^4-4 \bigl(92 \alpha ^{12}+794 \alpha ^{10}\\& \nonumber+2015 \alpha ^8+2420 \alpha ^6+1558 \alpha ^4+529 \alpha ^2+71\bigr) \nu ^2-48 \bigl(\alpha ^2+1\bigr)^3
   \bigl(2 \alpha ^6-8 \alpha ^4-3 \alpha ^2-1\bigr)\bigr) z^6\\& \nonumber+\nu  \bigl(-3 \bigl(\alpha ^3+\alpha \bigr)^2 \bigl(6 \alpha ^6-25 \alpha ^4-16 \alpha
   ^2-1\bigr) \nu ^6-2 \bigl(9 \alpha ^{12}+26 \alpha ^{10}+200 \alpha ^8+234 \alpha ^6+69 \alpha ^4\\& \nonumber+88 \alpha ^2+10\bigr) \nu ^4+4 \bigl(64 \alpha
   ^{12}+314 \alpha ^{10}+585 \alpha ^8+514 \alpha ^6+225 \alpha ^4+57 \alpha ^2+5\bigr) \nu ^2+16 \bigl(\alpha ^2\\& \nonumber+1\bigr) \bigl(10 \alpha ^{10}-24
   \alpha ^8-94 \alpha ^6-96 \alpha ^4-44 \alpha ^2-7\bigr)\bigr) z^5-\bigl(\alpha ^2 \bigl(\alpha ^2+1\bigr) \bigl(18 \alpha ^8+33 \alpha ^6-153 \alpha
   ^4\\& \nonumber-143 \alpha ^2-7\bigr) \nu ^6+2 \bigl(41 \alpha ^{12}+152 \alpha ^{10}+235 \alpha ^8+253 \alpha ^6+76 \alpha ^4+27 \alpha ^2+2\bigr) \nu ^4+4
   \bigl(\alpha ^2\\& \nonumber+1\bigr) \bigl(18 \alpha ^{10}-53 \alpha ^8-154 \alpha ^6-122 \alpha ^4-30 \alpha ^2-1\bigr) \nu ^2-16 \bigl(\alpha ^2+1\bigr)^4
   \bigl(2 \alpha ^4-4 \alpha ^2-1\bigr)\bigr) z^4\\& \nonumber+2 \alpha ^2 \nu  \bigl(\bigl(3 \alpha ^{10}+16 \alpha ^6+62 \alpha ^4+61 \alpha ^2+2\bigr) \nu ^4-2
   \bigl(\alpha ^2+1\bigr) \bigl(\alpha ^8+19 \alpha ^6+31 \alpha ^4+24 \alpha ^2+3\bigr) \nu ^2\\& \nonumber-4 \bigl(\alpha ^2+1\bigr)^2 \bigl(5 \alpha ^2
   \bigl(\alpha ^4-2\bigr)-4\bigr)\bigr) z^3+2 \alpha ^2 \nu ^2 \bigl(-3 \bigl(\alpha ^2-1\bigr) \bigl(\alpha ^3+\alpha \bigr)^2 \nu ^4+\alpha ^2
   \bigl(3 \alpha ^8+4 \alpha ^6\\& \nonumber+4 \alpha ^4+8 \alpha ^2+17\bigr) \nu ^2+2 \bigl(\alpha ^2+1\bigr)^2 \bigl(3 \alpha ^6-2 \alpha ^4-3 \alpha
   ^2-1\bigr)\bigr) z^2+4 \alpha ^4 \bigl(\alpha ^4-1\bigr) \nu ^3 \bigl(\alpha ^2\\& \nonumber-2 \nu ^2+1\bigr) z-4 \alpha ^4 \bigl(\alpha ^4-1\bigr) \nu
   ^4\bigr).
\end{align}

$C_{V,p}$ is given by

\begin{equation}
     C_{V,p}(\nu,z,\alpha)=\frac{{\cal N}_\text{6}(\nu, z,\alpha)}{{\cal D}_\text{6}(\nu, z,\alpha)},
\end{equation}

\noindent where

\begin{equation}
    \begin{split}
        \label{CVpnumerator}
        {\cal N}_\text{6}(\nu, z,\alpha)\equiv &\pi \ell_3(\nu)  \nu  \sqrt{\nu ^2+1} z \bigl(\alpha ^2 \bigl(z^2-1\bigr)-1\bigr)^2 (\nu  z+1) \bigl(\nu  \bigl(z^2-1\bigr)+2 z\bigr) \bigl(-4 \alpha ^2 z^2\\&+\nu  z
   \bigl(\alpha ^2 \bigl(z^4+3\bigr)+z^2+3\bigr)+2\bigr) \bigl(z^2 (\nu  z+1)+\alpha ^2 \bigl(-z^4+2 \nu  z^3-1\bigr)\bigr)
    \end{split}
\end{equation}

\noindent and

\begin{equation}
    \begin{split}
        \label{CVpdenominator}
        {\cal D}_\text{6}(\nu, z,\alpha)\equiv &\bigl(-\bigl(z^2 (2 \nu  z+3)\bigr)+\alpha ^2 \bigl(3 z^4-4 \nu  z^3+1\bigr)-1\bigr) \bigl(-z^2 (\nu  z+1) \bigl(4 \nu ^3 \bigl(12 z^5+z\bigr)+4 \nu  z
   \bigl(17 z^2-8\bigr)\\&+24 z^2+\nu ^4 z^2 \bigl(13 z^4+8 z^2+3\bigr)+\nu ^2 \bigl(80 z^4-30 z^2+2\bigr)-8\bigr)+\alpha ^2 \bigl(8 z^2 \bigl(9 z^4-8
   z^2+3\bigr)\\&+4 \nu  z \bigl(61 z^6-73 z^4+30 z^2-2\bigr)+\nu ^4 z^2 \bigl(77 z^8-190 z^6+24 z^4-10 z^2+3\bigr)+2 \nu ^3 z \bigl(118 z^8-201 z^6\\&+90
   z^4-5 z^2+2\bigr)+2 \nu ^2 \bigl(170 z^8-241 z^6+117 z^4-7 z^2+1\bigr)+2 \nu ^5 z^5 \bigl(3 z^6-26 z^4-15 z^2-6\bigr)\bigr)\\&-\alpha ^4 \bigl(8 (z-1)
   z^2 (z+1) \bigl(9 z^4-4 z^2+3\bigr)+2 \nu ^2 (z-1) (z+1) \bigl(118 z^8-195 z^6+109 z^4-13 z^2+1\bigr)\\&+4 \nu  z \bigl(53 z^8-100 z^6+84 z^4-28
   z^2+3\bigr)+\nu ^4 z^2 \bigl(17 z^{10}-202 z^8+186 z^6-88 z^4-3 z^2-6\bigr)+2 \nu ^3 z \\& \times\bigl(58 z^{10}-243 z^8+257 z^6-107 z^4+13 z^2-2\bigr)+\nu ^5
   z^5 \bigl(z^8-18 z^6+64 z^4+34 z^2+15\bigr)\bigr)+\alpha ^6 z\\& \times \bigl(8 z \bigl(z^2-1\bigr)^2 \bigl(3 z^4+1\bigr)-2 \nu ^5 z^4 \bigl(z^8-6 z^6+12
   z^4+6 z^2+3\bigr)+4 \nu  (z-1) (z+1) \bigl(15 z^8-22 z^6\\&+24 z^4-6 z^2+1\bigr)+\nu ^2 \bigl(44 z^{11}-206 z^9+328 z^7-220 z^5+92 z^3-6 z\bigr)+2 \nu
   ^3 z^2 \bigl(4 z^{10}-71 z^8\\&+118 z^6-120 z^4+30 z^2-9\bigr)+\nu ^4 z \bigl(z^{12}-26 z^{10}+121 z^8-60 z^6+51 z^4+6 z^2+3\bigr)\bigr)\bigr).
    \end{split}
\end{equation}

\end{widetext}

\bibliographystyle{apsrev4-1}
\bibliography{citations}

\end{document}